\documentclass[trackchanges, twocolumn, twocolappendix]{aastex701}
\usepackage{amsmath}
\usepackage{amssymb}
\usepackage{comment}

\newcommand{\Teff}{\mathrm{T}_\mathrm{eff}}
\newcommand{\logTeff}{\log{(\Teff)}}
\newcommand{\logg}{\log(g)}
\newcommand{\FeH}{\mathrm{[Fe/H]}}
\newcommand{\FeHi}{\mathrm{[Fe/H]}_\mathrm{i}}
\newcommand{\FeHs}{\mathrm{[Fe/H]}_\mathrm{s}}
\newcommand{\aFe}{[\alpha/\mathrm{Fe]}}
\newcommand{\aFei}{[\alpha/\mathrm{Fe]}_\mathrm{i}}
\newcommand{\aFes}{[\alpha/\mathrm{Fe]}_\mathrm{s}}

\newcommand{\Mi}{M_\mathrm{init}}

\newcommand{\amlt}{\alpha_\mathrm{MLT}}
\newcommand{\solaramlt}{\alpha_{\mathrm{MLT},\odot}}
\newcommand{\amltA}{\alpha_\mathrm{A}}
\newcommand{\amltB}{\alpha_\mathrm{B}}

\newcommand{\Ro}{\mathrm{Ro}}

\newcommand{\scaleheight}{H_P}

\newcommand{\dydz}{\frac{\Delta Y}{\Delta Z}}
\newcommand{\fovcore}{f_\mathrm{ov,core}}

\newcommand{\deltanu}{\Delta\nu}
\newcommand{\numax}{\nu_\mathrm{max}}

\newcommand{\Omegasync}{\Omega_\mathrm{sync}}
\newcommand{\Omegacrit}{\Omega_\mathrm{crit}}

\newcommand{\debs}{DEBs}
\newcommand{\interferometric}{interferometric}

\newcommand{\mistone}{MIST v.1}
\newcommand{\misttwo}{MIST v.2}
\newcommand{\misty}{\texttt{MISTy}}
\newcommand{\Ndebcat}{370}

\shorttitle{Classical Cases of Non-Solar Mixing Length}
\shortauthors{Woody et al.}

\graphicspath{{./}}
\begin{document}

\title{Classical Cases of Non-Solar Mixing Length in Low Mass Stars}

\correspondingauthor{Rebecca Woody}
\email{rebecca.woody@cfa.harvard.edu}

\author[0000-0002-0721-6715]{Rebecca Woody}
\affiliation{Center for Astrophysics $|$ Harvard \& Smithsonian, 60 Garden Street, Cambridge, MA 02138, USA}
\email{rebecca.woody@cfa.harvard.edu}

\author[0000-0002-1590-8551]{Charlie Conroy}
\affiliation{Center for Astrophysics $|$ Harvard \& Smithsonian, 60 Garden Street, Cambridge, MA 02138, USA}
\email{cconroy@cfa.harvard.edu}

\author[0000-0002-1617-8917]{Phillip A. Cargile}
\affiliation{Center for Astrophysics $|$ Harvard \& Smithsonian, 60 Garden Street, Cambridge, MA 02138, USA}
\email{pcargile@cfa.harvard.edu}

\author[0000-0002-4442-5700]{Aaron Dotter}
\affiliation{Department of Physics and Astronomy, Dartmouth College, 6127 Wilder Laboratory, Hanover, NH 03755, USA}
\email{aaron.l.dotter@dartmouth.edu}

\begin{abstract}
\noindent The treatment of convection is one of the largest remaining uncertainties in models of low mass stars.  The Mixing Length Theory (MLT) for super-adiabatic convection has come under particular scrutiny.  Evidence is mounting that a single value of the mixing length parameter ($\amlt$) should not be applied universally across stellar model grids, although simulations and asteroseismic calibrations disagree on how it should vary with stellar parameters.  A variable $\amlt$ causes critical systematic changes to the ages of low mass stars, the key age tracers for galactic and exoplanetary studies.  In this work we use classical (non-seismic) geometric radii and specialized MIST isochrones with a variable $\amlt$ to measure mixing length in detached eclipsing binaries (DEBs) and interferometrically resolved stars, finding plenty of evidence for non-solar $\amlt$.  In particular we find that standard evolutionary models overpredict the radii of metal-poor interferometric stars and that a sub-solar $\amlt/\solaramlt\sim0.50-0.75$ can neatly explain this discrepancy, consistent with other empirical calibrations and in stark contrast to simulations.  We infer low convective efficiency in cool, rapidly rotating lower main-sequence stars, consistent with the known phenomenon of magnetically-induced radius inflation.  We also identify several slowly rotating DEBs whose measurements cannot be reproduced by standard isochrones, and we propose these systems as useful case studies of non-solar mixing length in stars outside of the Sun.  However, we find no definitive population level correlations between $\amlt$ and stellar parameters among the DEBs, calling into question the existence of a ubiquitous mixing length calibration in low mass stars.
\end{abstract}

\section{Introduction}
\label{section:introduction}
Low mass stars are numerous and long lived, making them the ideal age tracers to understand the evolutionary history of exoplanetary systems \citep{SilvaAguirre_2015, Bixel_2020, Unterborn_2022, Ware_2026}, the Milky Way at large \citep{Haywood_2013, Gallart_2019, Bonaca_2020, Xiang_2022, Conroy_2022, Woody_2025}, and stellar populations in other galaxies \citep{Mateo_1998, Tolstoy_2009, Weisz_2014, Simon_2019}.  Stellar models have been remarkably successful at reproducing most observations of low mass stars for decades now, yet stellar ages in particular are still notoriously difficult to measure reliably, and as the quality of the data has continued to improve the remaining uncertainties in the models are limiting further progress.  Improvements in stellar models have comparatively not kept up with the improvements in data and so we have reached the point where the models are systematically limiting absolute stellar ages to accuracies of $\sim20\%$ \citep{Tayar_2022, Joyce_2023a, Li_2024, Tayar_2025}.  

Some of the largest remaining uncertainties in models of low mass stars involve the treatment of convection.  Convection is inherently a 3D turbulent advective process that must be approximated by simpler prescriptions in 1D evolutionary models.  The treatment of super-adiabatic convection in stellar envelopes in particular has come under recent scrutiny.  The standard formalism for such convection is Mixing Length Theory \citep[MLT;][]{Vitense_1953, Bohm-Vitense_1958}, wherein convection is approximated as a diffusive process with a single characteristic transport length scale $\lambda_\mathrm{MLT}=\amlt \scaleheight$, where $\scaleheight$ is the local pressure scale height and $\amlt$ is the so called ``mixing length parameter", a numerical parameter of order unity.  This parameter governs the efficiency of convection and the adiabatic stratification of the envelope, thereby setting the radius and surface temperature of the model.  Other formulations for convection in 1D exist \citep[e.g., full spectrum turbulence;][]{Canuto_1991, Canuto_1992, Canuto_1996}, but MLT remains the default within most general purpose isochrone libraries \citep[e.g., Y$^2$, BaSTI, DSEP, PARSEC, MIST;][]{Yi_2001, Pietrinferni_2004, Dotter_2008, Bressan_2012, Choi_2016}.

The value of the mixing length parameter cannot be predicted from first principles and must instead be calibrated by either observations or simulations.  In most cases a clean calibration of this parameter using observations is not possible for the following reasons.  A stellar model is effectively a map between the inputs of mass, composition in the form of helium and metal mass fractions $Y$ and $Z$, age, and mixing length to outputs like radius and temperature, conditioned on the assumptions of input physics (e.g., opacities, equation of state).  For a typical field star we might have observations constraining the star's temperature, luminosity via distance, mass via surface gravity, and metallicity.  This is four constraints against the five input parameters, and so there exists a 1D locus of stellar models in $Y-$age$-\amlt$ space that can match these observations.  Even with the common assumption of a fixed helium to metal enrichment ratio $\dydz$ there persists a degeneracy between age and $\amlt$ that cannot be broken without some other independent constraint on age.  

This age$-\amlt$ degeneracy is a fundamental difficulty in calibrating $\amlt$ because the Sun is the only star with an age known independently of stellar models, thanks to our access to protoplanetary meteoritic material\footnote{Radioisotope dating of photospheric Th and U (nucleocosmochronology) is possible for a handful of stars \citep[e.g.,][]{Cowan_1991, Frebel_2008}.  However, this technique still permits uncertainties of order several Gyrs and is not precise enough for a useful $\amlt$ calibration.}.  Historically, the most common practice has been to calibrate a value of the mixing length for a solar model and to then apply that $\solaramlt$ to all other stars regardless of mass, composition, or evolutionary phase.  Though this simple assumption has proven remarkably successful in enabling models to faithfully reproduce stellar observables like temperature and luminosity, it has no real physical basis and any deviations from $\amlt = \solaramlt$ in reality will propagate through the age$-\amlt$ degeneracy to become troubling systematic biases in age inferences.  

\begin{figure*}[t!]
\epsscale{1.0}
\plotone{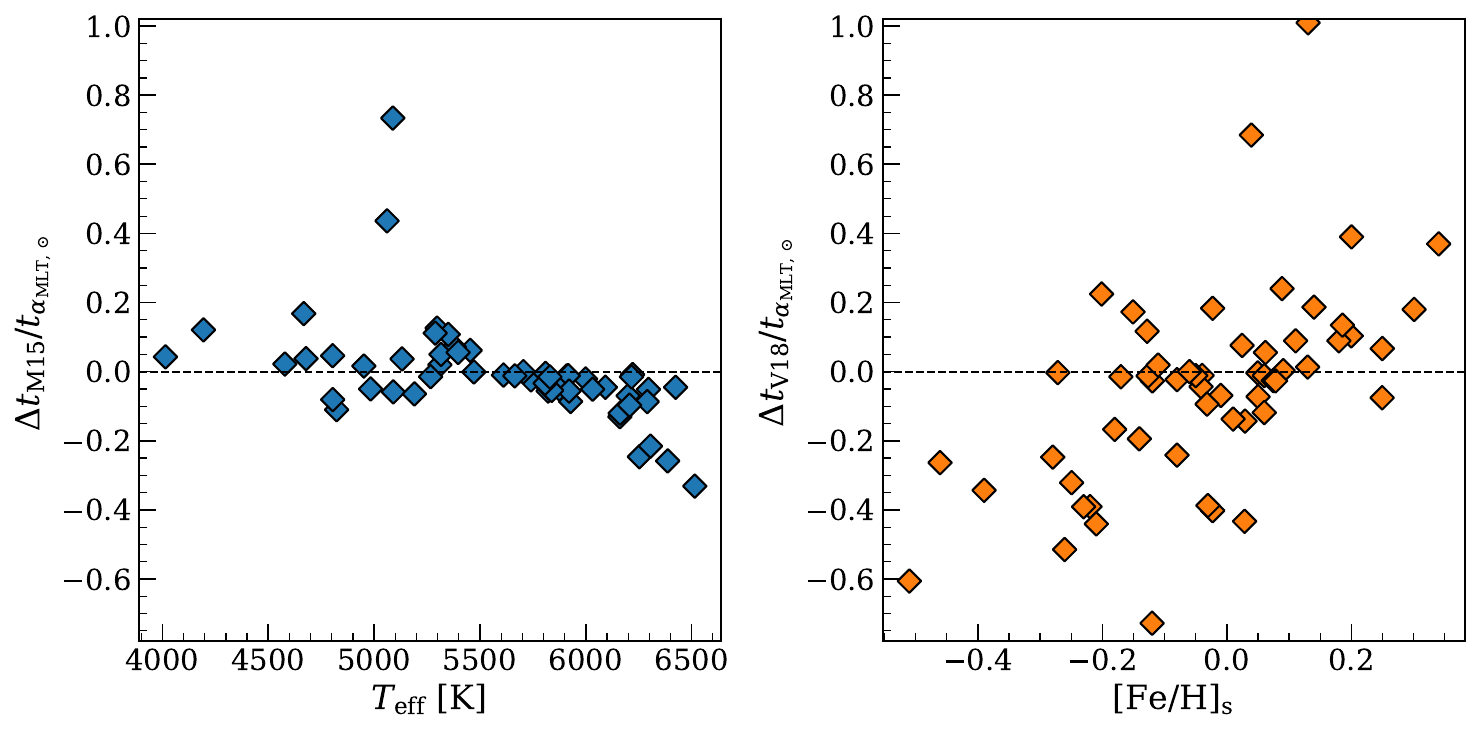}
\caption{Fractional difference in ages inferred for the \interferometric\ sample (Section \ref{subsection:irads}) when adopting either the solar-calibrated mixing length $\solaramlt=2.03$ or a mixing length from a literature calibration.
\emph{Left:} the \citet[M15;][]{Magic_2015} calibration based off the STAGGER 3D RHD simulations.  \emph{Right:} the \citet[V18;][]{Viani_2018} calibration fit to solar-like oscillators.  \citet{Magic_2015} predicts a mild dependence on $\logTeff$ and $\logg$ and only a weak dependence on $\FeH$.  \citet{Viani_2018} predicts a very strong dependence on $\FeH$, leading to a dramatic offset between the inferred ages of stars with non-solar metallicities.  
Both demonstrate that an incorrect choice of mixing length can systematically limit age inferences of Sun-like stars to an accuracy of $\sim 10 - 20\%$ and that this systematic only becomes more severe for stars increasingly dissimilar than the Sun.
\label{figure:irad_age_bias}}
\end{figure*}

An example of this age systematic can be seen in Figure \ref{figure:irad_age_bias} where we use our sample of interferometric stars (Section \ref{subsection:irads}) to compare two different non-solar mixing length prescriptions \citep{Magic_2015, Viani_2018} against the standard solar-calibrated assumption.  Stars increasingly dissimilar than the Sun are predicted to have larger deviations from $\solaramlt$ and thus will have larger biases in their ages.  Many stars in this sample are exoplanet hosts, highlighting how variations in mixing length can propagate into a variety of other areas of study that rely on low mass stellar models.  

In fact, radiative hydrodynamical (RHD) simulations of convection in stellar envelopes do suggest that $\amlt\neq\solaramlt$ generally, but that $\amlt$ should instead vary as a function of stellar surface properties \citep{Ludwig_1999, Freytag_1999, Trampedach_2013, Magic_2013, Magic_2015, Tanner_2013, Tanner_2013a, Tanner_2014, Tanner_2016}.  RHD simulations manifest convection self-consistently and therefore do not possess a ``mixing length" in the way that 1D models do, but a mapping can be made using the simulation's entropy structure.  Convection is efficient in the deep layers of a stellar envelope and so the temperature gradient is essentially adiabatic, i.e., the specific entropy is asymptotically constant.  The particular value of this asymptotic entropy depends on the choice of mixing length parameter in 1D models, while in simulations the entropy is an input variable that sets the adiabat of the hot, isentropic material rising from the bottom of the simulation boundary.  By matching this asymptotic entropy a correspondence can be made between a simulation and a 1D model of a particular mixing length.  Other definitions can be used to map the mixing length parameter onto simulations, e.g., the jump between the asymptotic entropy and the entropy minimum near the photosphere, or via the length scale of convective mass transport, but in all cases the derived mixing length behaves similarly \citep{Magic_2015}.

Under this mapping the mixing length parameter and the asymptotic entropy of the deep envelope's adiabat are anti-correlated.  Simulations like those of \citet{Ludwig_1999, Trampedach_2013, Magic_2015} and \citet{Tanner_2016} demonstrate that this entropy varies across the HR diagram, generally finding larger entropies and thus lower mixing lengths for stars with higher temperatures or lower surface gravities.  Some simulations have also explored the dependence on composition and found that at fixed HR position lower metallicities tend to result in lower entropy and higher mixing length, though this dependence is relatively weak \citep{Tanner_2013, Tanner_2013a, Magic_2013, Magic_2015}.  The magnitude of mixing length variation suggested by simulations is modest, $\sim\pm20\%$ relative to $\solaramlt$, but should still lead to evident discrepancies between sufficiently precise datasets and stellar models that assume $\amlt = \solaramlt$.

Puzzlingly, while empirical studies have found non-solar mixing lengths that correlate with stellar parameters and some find rough agreement with simulation predictions \citep{Joyce_2018a}, most of the trends are opposite to those suggested by the simulations.  \citet{Lebreton_2001} and \citet{Yildiz_2006} found that the slope of the main sequence $M-L$ relation in the Hyades requires a sub-solar mixing length at low masses.  \citet{Bonaca_2012} and \citet{Viani_2018} used an ensemble of Kepler targets with asteroseismic masses and radii to infer mixing lengths and then fit a trilinear relationship to $\logg$, $\logTeff$, and metallicity.  They found that mixing length on the main sequence decreases with both temperature and surface gravity, and that the sign of this correlation flips for more evolved stars on the lower red giant branch.  They also found a strong positive correlation between mixing length and  metallicity, in stark contrast to the simulations that suggest at most a weak anti-correlation with metallicity (see Table 7 in \citet{Viani_2018}). 

A positive correlation with metallicity has been reported elsewhere.  \citet{Tayar_2017} found that the range of temperatures spanned by giants in the second APOKASC catalog \citep{Pinsonneault_2018} was smaller than models predict at a fixed metallicity.  They found that the mismatched temperature scale could be neatly explained with a non-solar mixing length that increased $\Delta\amlt \sim 0.2/\mathrm{dex}$ in metallicity.  However, the RGB temperature scale is particularly sensitive to not only mixing length but also the choice of atmospheric boundary condition, and other studies have challenged this result \citep{Salaris_2018, Choi_2018a, Valle_2019}.  \citet{Joyce_2018} found that the very metal-poor subgiant HD 140283 ($\FeH\sim-2.5$) requires a mixing length $\amlt/\solaramlt \sim 0.5$ in order for models to reproduce its radius and temperature, though this result has also been questioned as an artifact of HD 140283's peculiar elemental abundance mixture \citep{Guillaume_2024, Lundkvist_2025}.  Still, \citet{Joyce_2018} go on to argue that the CMDs of other similarly metal-poor dwarfs and the globular cluster M92 further support the case for sub-solar mixing lengths at low metallicities.  

Empirical studies also tend to find variations in mixing length far greater in magnitude than simulations predict, often finding $\amlt\sim\pm50\%~\solaramlt$.  It is possible that this discrepancy arises from the other limitations of MLT; even if the 1D mixing length and resultant asymptotic entropy are calibrated to match simulations, MLT and RHD simulations predict qualitatively different superadiabatic gradients in the surface layers \citep{Magic_2015, Tanner_2016}.  There has been some recent work on a more sophisticated entropy calibration that accounts for this \citep{Magic_2016, Spada_2018, Spada_2019, Spada_2021, Manchon_2024}, but so far this method has not been widely adopted.

There is a clear need to move beyond a fixed, solar-calibrated mixing length, but there is still no consensus on the appropriate replacement.  In this paper we aim to test if either the asteroseismically calibrated or simulation predicted mixing length variations are supported by a set of independently constraining observations, namely well-measured geometric stellar radii.  We also identify a number of interesting systems that can serve as new case studies for testing non-solar mixing length.  In Section \ref{section:data} we define our sample of stars with classical, geometrically measured stellar radii and Section \ref{section:methods} outlines our procedure for inferring the mixing length parameter in these systems.  In Section \ref{section:debs_results} we discuss our findings with the detached eclipsing binaries, Section \ref{section:irad_results} discusses the interferometric stars, and Section \ref{section:cluster_results} briefly describes our results simultaneously fitting detached eclipsing binaries in clusters.  We discuss potential caveats to our work in Section \ref{section:caveats} and offer our conclusions in Section \ref{section:conclusion}.

\section{Data}
\label{section:data}
\subsection{Overview of Non-solar Calibrators}
The mixing length parameter primarily influences the radius and temperature of a stellar model, so to empirically constrain it requires stellar radii measured independently of evolutionary models.  There are three main ways this can be done.  Radial velocities and light curves of double-lined, detached eclipsing binaries (DEBs) can be solved to yield model independent, dynamical masses \& radii with a typical precision of $\sim 1\%$.  Some nearby bright stars can be resolved via interferometry and then their angular diameter can be combined with a parallax to yield a purely geometric measurement of linear radius, with the best reaching precisions of $\sim 1\%$.  Lastly, solar-like oscillations can be used to infer the sound speed structure of a star.  If only the global properties of the oscillation power spectrum ($\deltanu$, $\numax$) are measurable, then the stellar parameters can be inferred from scaling relations \citep{Brown_1991, Kjeldsen_1994, Bedding_2003}, which are accurate to $\sim2\%$ in radius and $\sim4\%$ in mass when including corrections for metallicity and evolutionary phase \citep{SilvaAguirre_2015, Sharma_2016, Aguirre_2017, Li_2020, Hon_2022, Li_2023}.  If individual mode frequencies can be identified then a more detailed asteroseismic modeling of the star's internal structure can be performed, yielding even more precise and robust mass and radius measurements \citep{Metcalfe_2012, Chaplin_2020, Brogaard_2022, Huber_2024, Hon_2024, Li_2025, Lundkvist_2025}.

The existing empirical trends between non-solar mixing length and stellar parameters focused on ensembles of global asteroseismic observables (\citealt{Bonaca_2012, Tayar_2017, Viani_2018, Li_2024}, but see \citealt{Valle_2019, Dennis_2026}), so in this work we instead focus on the first two scenarios where the radius is measured through purely geometric and classical (that is, non-asteroseismic) methods.  Though this sample will be smaller than the asteroseismic ensembles, purely geometric radii have and continue to be valuable independent tests on stellar models and non-solar mixing lengths in particular, with the fact that seismic scaling relations are validated against geometric radii underscoring their importance.  We therefore consider samples of FGK-type stars in detached eclipsing binaries or with an interferometrically resolved angular diameter, testing whether their observed radii support the proposed empirical or theoretical behaviors for a non-solar mixing length.  We also highlight several particularly interesting cases where a non-solar mixing length is clearly preferred to in order for models to match the observations.

\subsection{Eclipsing Binary Sample}
\label{subsection:debs}
Detached eclipsing binaries have long been used as one of the gold standards with which to test evolutionary models \citep{Andersen_1991, Torres_2010}.  Radial velocities measured from high-resolution echelle spectrographs together with high quality ground or space-based light curves allow for masses and radii to be measured to a precision of $\sim1\%$.  However, because of the relatively high resource commitment to obtain the measurements needed to fully solve a system's orbit, only a small number of all known eclipsing systems have been characterized to this level of precision.  The Detached Eclipsing Binary Catalog\footnote{https://www.astro.keele.ac.uk/jkt/debcat/} \citep[DEBCat;][]{Southworth_2014} keeps an up to date record of virtually all of such systems in the literature, making it a superset of the \citet{Torres_2010} catalog.  At the time of writing it contains \Ndebcat\ systems, including binaries of all main sequence spectral types and several giant-giant binaries of low and intermediate mass stars.

Starting from the March 23, 2026 version of DEBCat we select systems where both components have $0.5M_\odot<M<1.8M_\odot$, both have masses and radii measured to a precision better than 2\%, at least one component has $\Teff<6800$ K, and where a literature estimate of their metallicity is available.  This mass range focuses on stars that are most relevant for exoplanetary and Galactic archaeological studies, and the temperature range ensures that at least one star in the system is sensitive to the choice of mixing length.  Our ultimate aim is to isolate any signal of a non-solar mixing length, so we further select systems where the effects of other uncertain and confounding stellar modeling prescriptions can be avoided.  We require $\logg>1.0$ for both binary components, beyond which cumulative mass loss on the RGB begins to dominate over the measurement uncertainty in the dynamical masses.  We also remove any pre-main sequence binaries and any systems containing a helium burning giant.

Another concern is rotation, which alters the structure and lifetime of a star through additional mixing and in the extreme cases deformation.  Low mass stars ($M \lesssim 1.2 M_\odot$) have large convective envelopes that drive efficient magnetic braking, quickly slowing down any initial rotation \citep{Schatzman_1962, Kraft_1967, Skumanich_1972}.  However, stars in eclipsing binaries are often rotationally synchronized to their orbital period, meaning that binary low mass stars often rotate more quickly than their non-binary counterparts.  The situation is somewhat reversed for intermediate mass stars ($M\gtrsim1.2 M_\odot$) which have thin or completely lack convective envelopes and therefore keep their higher initial rotation rates.  In these cases synchronization can instead cause a star to be rotating more slowly than is typical for their field counterparts.  The vast majority of systems in DEBCat have relatively slow rotation rates even if assuming tidal synchronization in all cases.  Nevertheless we only keep systems where both components have $\Omegasync < 0.15\Omegacrit$, with $\Omegasync = \frac{2 \pi}{P_\mathrm{orbit}}$ and  $\Omegacrit = \sqrt{\frac{GM}{R^3}}$.  This is roughly when the evolution of rotating and non-rotating \misttwo\ models begin to differ significantly \citep[see also][]{Fritzewski_2024}.  This way we keep only those systems that are safely representative of single star evolution.

To this sample we add the stars $\alpha$ Centauri A \& B, Procyon A, and $\mu$ Cassiopeiae A.  These are members of visual binaries that have radii measured through interferometry instead of eclipses but who otherwise satisfy the above criteria, with dynamical masses and geometric radii measured to better than $2\%$ \citep{Kervella_2016, Kervella_2017, Bond_2015, Bond_2020}.  The companion Procyon B is a white dwarf and $\mu$ Cassiopeiae B is a low mass M dwarf ($0.17M_\odot$) so we do not consider them in our analysis.

Several of the DEBs in our sample are known members of open or globular clusters.  The usual assumption that clusters are simple stellar populations offers a particularly strong joint age constraint to break the degeneracy with $\amlt$, making clusters excellent testing grounds for the relative dependence of mixing length on stellar parameters, particularly mass \citep{Yildiz_2006}.  The clusters represented in our sample are the open clusters NGC 6791 and Ruprecht 147, and the globular clusters 47 Tuc, M4, M55, NGC 3201, and NGC 6362.  M55 only contains a single binary in DEBCat which did not strictly pass our original selection criteria \citep[V54;][]{Kaluzny_2014}, but we elect to include it anyhow because it is the most metal-poor entry in the entirety of DEBCat, with a metallicity of $\FeH\approx-1.9$.  Table \ref{table:clusters} briefly lists the clusters and their member binaries that we consider in this work.  

Our final sample contains 184 stars across 93 binary systems with precise masses \& radii.  Figure \ref{figure:data_and_model_grid_coverage} plots these stars in mass$-$metallicity space and in a $\Teff-$Radius HR diagram colored by their metallicities, along with the sample of interferometric radius stars we define below.  The globular cluster DEBs extend the metallicity range of the sample considerably, enabling a strong test of the proposed dependence of mixing length on metallicity.

\begin{deluxetable}{lll}
\tablecaption{Cluster DEBs\label{table:clusters}}
\tablehead{
\colhead{Cluster} & \colhead{DEB Members} & \colhead{References}}
\startdata
NGC 6791 & V565 Lyr, V568 Lyr & \citet{Brogaard_2010, Brogaard_2012} \\
Ruprecht 147 & TYC 6296-96-1, TYC 6296-2012-1 & \citet{Torres_2018, Torres_2019} \\
 & EPIC 219552514 & \citet{Torres_2020, Torres_2021} \\
\hline
47 Tuc & V69, E32 & \citet{Thompson_2020} \\  
M4 & V66, V69 & \citet{Kaluzny_2013} \\
M55 & V54 & \citet{Kaluzny_2014} \\
NGC 3201 & V138, V139, V141 & \citet{Rozyczka_2022} \\
NGC 6362 & V40, V41 & \citet{Kaluzny_2015} 
\enddata
\tablecomments{Open and globular clusters represented in our \debs\ sample, their member detached eclipsing binaries, and the literature sources for their masses, radii, temperatures, and metallicities.}
\end{deluxetable}

\subsection{Interferometric Sample}
\label{subsection:irads}
We adopt the third version of the \emph{Gaia} FGK benchmark star (GBSv3) catalog \citep{Blanco-Cuaresma_2014, Soubiran_2024, Casamiquela_2026} as the parent catalog for our sample of interferometric radii stars.  Their goal was to build a catalog of stars to serve as benchmarks for stellar spectroscopy and abundance analyses, and they therefore paid particular attention to covering the full range of stellar atmospheric parameter space.  To that end they supplemented the sample of \citet{Salsi_2020} with additional metal-poor targets that have had recent interferometric measurements \citep{Karovicova_2020, Karovicova_2022a}, yielding a final catalog of 192 stars with reliable angular diameters.  All of these stars have good parallax measurements and therefore have purely geometric measurements of their linear radii.

From this catalog we select stars using similar criteria as was used for the \debs\ sample in Section \ref{subsection:debs}.  First, we restrict to stars whose measured linear radii are precise to $\leq2\%$ and have $\Teff < 6800$ K.  We restrict to $R < 8R_\odot$ in order to avoid the ambiguity between first ascent RGB and helium burning stars.  However, we choose to include three exceptions that exceed this radius cut: HIP 27530 ($\gamma$ Pic), HIP 69673 (Arcturus, $\alpha$ Boo), and HIP 92167.  These three stars have asteroseismic measurements that confirm their evolutionary status as first ascent red giants \citep{Hon_2022, Tarrant_2007, Mosser_2014}.

\begin{figure*}[t!]
\epsscale{1.0}
\plotone{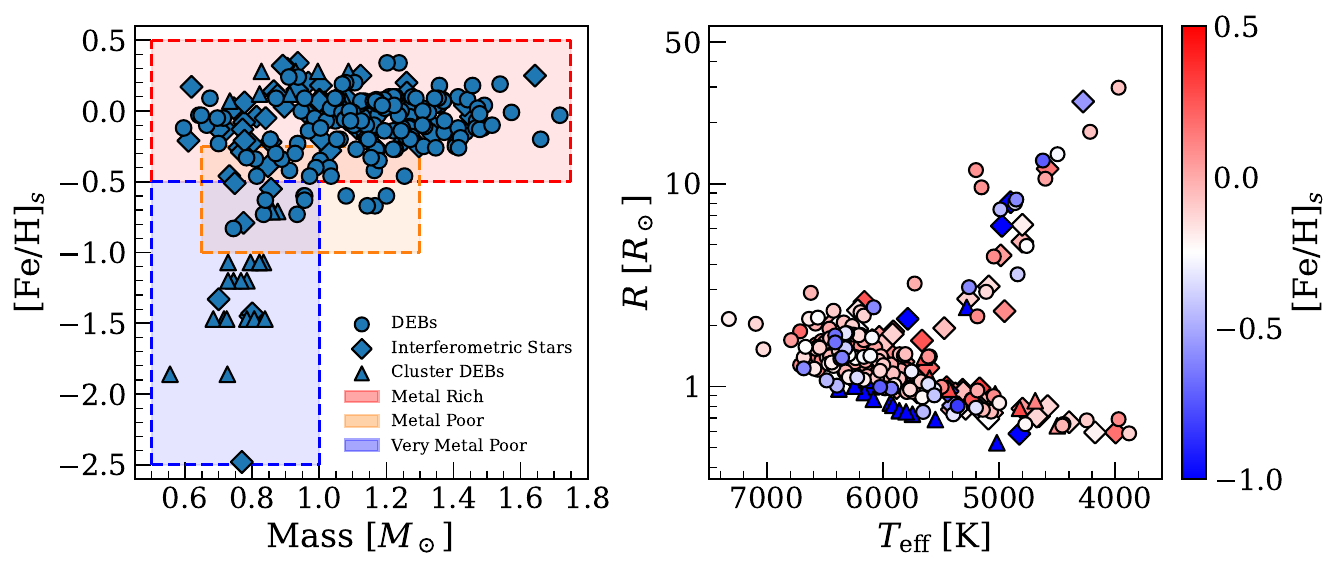}
\caption{\emph{Left:} Our detached eclipsing binary (\debs) and interferometric radii star samples in mass$-$metallicity space, plotted as circles and diamonds respectively.  The parameter space coverage of our three model grids are plotted as shaded regions.  \emph{Right:} The samples plotted in the $\Teff-$Radius diagram, colored by metallicity.  Typical temperature uncertainties are 100 K and 40 K for the DEBs and interferometric stars respectively.  The sample covers the full extent of FGK-type main-sequence stars, contains about two dozen giant stars, and spans nearly three dex in metallicity, making it suited for testing the mixing length trends suggested by RHD simulations and ensemble asteroseismology.  \label{figure:data_and_model_grid_coverage}}
\end{figure*}

The stars in this sample do not have model-independent masses, but \citet{Soubiran_2024} do provide mass estimates based on the BaSTI \citep{Pietrinferni_2004, Pietrinferni_2006} and STAREVOLVE \citep{Lagarde_2012, Lagarde_2017} evolutionary track libraries.  We use these masses as a guide to define our sample, but we do not use them during the actual mixing length inference in Section \ref{subsection:fitting}.  We make a conservative cut and only include stars with $M_\mathrm{BaSTI} + 3\sigma_{M_\mathrm{BaSTI}} < 1.8M_\odot$.  Lastly, $\alpha$ Centauri A \& B, Procyon A, and $\mu$ Cassiopeiae A are part of the GBSv3 catalog and are retained by our cuts, but we remove them here and instead consider them as part of the \debs\ sample because of their dynamical masses.  Applying all of these cuts leaves us with 69 stars in our \interferometric\ sample, plotted in Figure \ref{figure:data_and_model_grid_coverage} as diamonds.

\section{Methods}
\label{section:methods}
\subsection{Context of $\amlt$ in Stellar Models}
The mixing length parameter is a \emph{numerical} parameter in the sense that it does not directly correspond to any observable physical quantity of a star (the entropy mapping is only approximate and the entropy of the deep adiabat is not directly observable).  It absorbs physical uncertainties and numerical artifacts of the evolutionary code, meaning that the exact value of the mixing length parameter will not be consistent across different codes nor even within the same code run under different physical inputs (e.g., opacities, atmospheric boundary conditions).  Any test of the mixing length parameter must therefore be done in a differential sense, comparing to the solar calibrated value particular to that physical and numerical configuration.  Thus for the rest of the manuscript all mixing length parameters are reported relative to our solar-calibrated value of $\solaramlt=2.03$ unless otherwise stated.  

Another critical note to keep in mind is that any inferred non-solar mixing length, at a fundamental level, only implies that our stellar models incorrectly predict the radius and temperature of a star at a particular mass, composition, and age.  There are other physical prescriptions and approximations in 1D stellar modeling that could instead be the culprit behind a mismatch between models and observations.  In Section \ref{section:caveats} we discuss some of these other model uncertainties.

\subsection{Model Grid}
\label{subsection:model_grid}
We base our model grid off of the \misttwo\ library of stellar tracks and isochrones \citep{Choi_2016, Dotter_2026}, rebuilding a portion of the library to include mixing length as a variable.  Here we'll briefly summarize the physical prescriptions and changes in v.2 that are relevant to this study.  

The \misttwo\ grid is constructed using revision 11701 of MESA \citep{Paxton_2010, Paxton_2013, Paxton_2015, Paxton_2018, Paxton_2019}.  It adopts the solar abundance mixture of \citet[][(GS98)]{Grevesse_1998} instead of the \citet{Asplund_2009} mixture used in \mistone, motivated by recent revisions of the solar metallicity back up towards that of GS98 \citep{Villante_2014, Caffau_2010, Magg_2022}.  Under this new scale the solar-calibrated linear helium enrichment law has a slope $\dydz = 1.32$.  In addition to the usual solar scaled mixtures, \misttwo\ also allows for variations in the $\alpha$ elements, with the abundances of O, Ne, Mg, Si, S, Ar, Ca, and Ti relative to Fe varied in lock step via the parameter $\aFe$.  Increasing $\aFe$ increases total metallicity [M/H] relative to $\FeH$.  

\misttwo\ adopts the scheme of \citet{Dotter_2022} to define the atmospheric boundary condition by incorporating convection when integrating the $T(\tau)$ relation.  This removes the ambiguity in the $\Teff$ scale introduced by the arbitrary choice of depth at which the boundary condition is placed \citep{Choi_2018a}, a change that is especially relevant for the large convective envelopes of giant stars.  This new scheme does however introduce a small internal inconsistency to the present work:  the boundary condition tables were built using \misttwo's solar-calibrated mixing length of $\solaramlt = 2.03$.  Here we test for non-solar mixing lengths in the \emph{interior} model calculated with MESA, but we do not rebuild the entire atmosphere tables with these non-solar mixing lengths.  We have confirmed that this inconsistency is indeed small, leading to at worst a systematic bias of $\Teff \lesssim10$ K, which is subdominant to any realistic observational uncertainties in $\Teff$.

\misttwo\ by default only contains rotating models, which follow the same scheme for rotation as \mistone\ where rotation is initialized with a rate $\Omega_\mathrm{ZAMS}$ at the zero age main sequence (ZAMS), with $\Omega_\mathrm{ZAMS}=0.0$ for $M\leq1.2M\odot$ and then smoothly ramping up until $\Omega_\mathrm{ZAMS}=0.4\Omega_\mathrm{crit}$ at $M\geq1.8M_\odot$.  In Section \ref{section:data} we specifically selected systems whose rotation was slow enough as to not cause significant changes to structure and evolution, so we build our grid using only non-rotating models.

To efficiently cover all of our sample we split up parameter space into a three different metallicity sections.  The metal-rich grid is built over masses $0.50M_\odot$ to $1.75M_\odot$, $-0.5 \leq \FeH \leq +0.5$ with a $0.25$ dex spacing, $\aFe=0.0$, and over nine values of mixing length parameter (1.0, 1.3, 1.6, 1.8, 2.0, 2.2, 2.4, 2.7, 3.0).  Note that these are the absolute not relative values for the mixing length parameter, ranging from $50\%$ to $150\%$ $\solaramlt$.  The masses are built with a spacing of $0.05M_\odot$ and with additional grid points at $1.175M_\odot,~1.225M_\odot$, and $1.275M_\odot$ to help resolve the onset of the convective hook.

The metal-poor grid is built in a similar fashion.  It covers $0.65M_\odot\leq M \leq1.30 M_\odot$ with the same spacing and again including the additional masses near the onset of the hook, metallicities $-1.00 \leq \FeHi \leq -0.25$, $\aFe=+0.2$, and over the same set of $\amlt$ values.  Lastly the very metal-poor grid is built over masses $0.50M_\odot\leq M \leq1.00 M_\odot$, $-2.50 \leq \FeHi \leq -0.5$, $\aFe=+0.4$, and again over the same $\amlt$ values.  Figure \ref{figure:data_and_model_grid_coverage} shows the coverage in mass$-$metallicity space alongside the extent of our data samples.

The majority of our \debs\ sample do not have literature measurements of $\aFe$ available so we chose to progressively increase $\aFe$ as grid metallicity decreases as a way to roughly capture the observed $\FeH-\aFe$ pattern of the Milky Way's thick disk \citep{Haywood_2013}.  We intentionally built the grids with some partial overlap at $\FeH\approx-0.50$, and later in Section \ref{section:caveats} we use this overlap to test how sensitive our mixing length inferences are to an incorrectly assumed $\aFe$.

All tracks are initialized on the pre-main sequence as is typical in MESA and MIST, with the metal-rich and metal-poor models evolved until reaching $\logg=1.0$ on the red giant branch, while the very metal-poor models were evolved until $R = 15R_\odot$.  We transform the tracks to equal evolutionary points \citep[EEP;][]{Dotter_2016} but we do not use the default spacing in MIST and instead adopt a spacing tailored for our purposes, placing 300 EEPs between the ZAMS and the track's terminus.  We have confirmed that this spacing is sufficient to resolve all of the fine details of the evolutionary tracks including the convective hook and the red giant branch bump. 

\begin{figure}[t!]
\epsscale{1.0}
\plotone{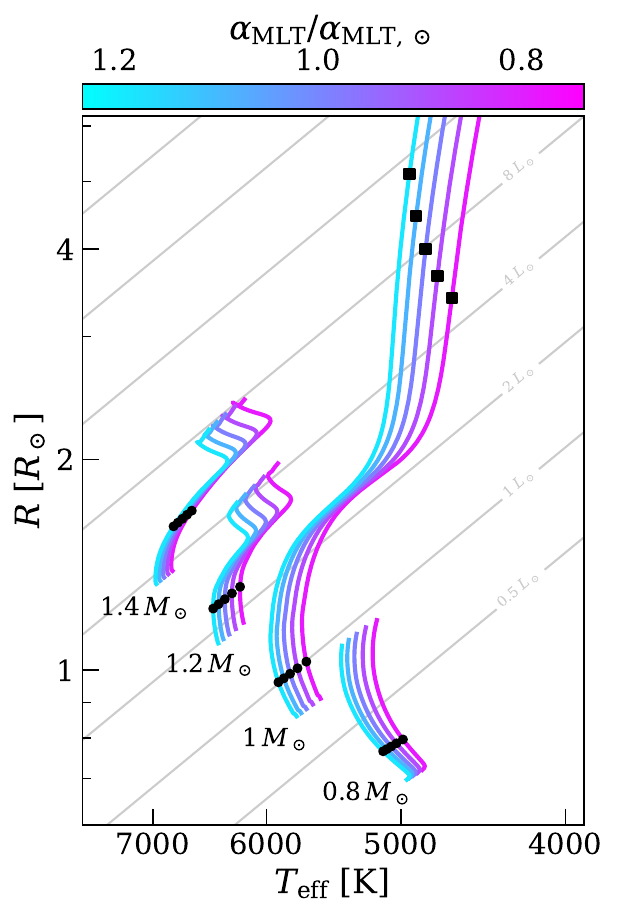}
\caption{Tracks of $0.8, ~1.0, ~1.2, ~\mathrm{and}~1.4~M_\odot$ solar metallicity models with different values of $\amlt$. Tracks with $M \neq 1.0M_\odot$ are truncated at the terminal age main sequence for visual clarity.  The dots mark points of equal age within each set of tracks.  Low mass stars have a small super-adiabatic jump in the outermost layers and so changing $\amlt$ has only a mild effect.  Sensitivity to $\amlt$ increases with mass until roughly $\sim 1.2 M_\odot$, where the shrinking size of the convective envelope then makes higher masses less sensitive to $\amlt$ on the main sequence.  Varying $\amlt$ at fixed age on the main sequence preserves luminosity, but post-main sequence phases are very sensitive to the choice of $\amlt$.  \label{figure:amlt_effect}}
\end{figure}

Examples of tracks with different values of $\amlt/\solaramlt$ are plotted in Figure \ref{figure:amlt_effect}.  A larger mixing length increases convective efficiency, decreasing the model's radius.  The luminosity and therefore lifetimes of main-sequence stars are largely insensitive to $\amlt$, and so the change in radius and surface temperature are linked via the Stefan-Boltzmann law.  Sensitivity to $\amlt$ on the main sequence is strongest at roughly solar temperatures.  Sensitivity decreases at high temperatures due to the vanishing convective envelope and decreases at low temperatures due to a smaller super-adiabatic jump in the outermost layers.  Post-main sequence phases have deep, cool convective envelopes and are therefore very sensitive to $\amlt$.

\subsection{\misty\ NN Emulator}
\label{subsection:misty_nn}
\begin{figure}[t!]
\epsscale{1.0}
\plotone{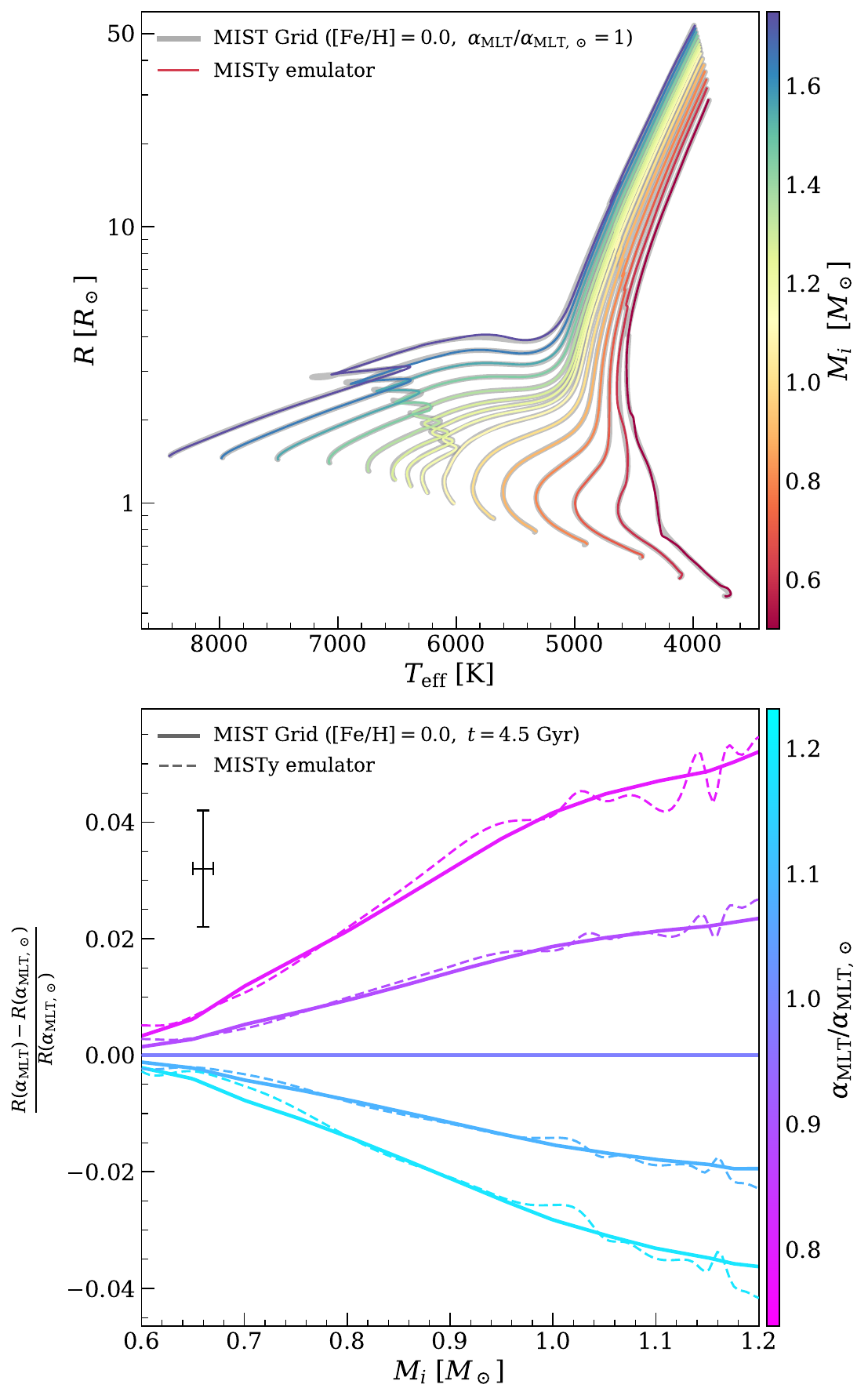}
\caption{Stellar tracks from the metal-rich model grid compared to MISTy NN emulated tracks.  A typical error bar of 1\% in mass and radius is plotted for reference.  The emulation is accurate and contributes a small but subdominant numerical uncertainty to our $\amlt$ inferences.  The emulation is particularly accurate for phases away from the convective hook.  The emulation residuals are unbiased and their 95th percentiles are $12$ K in $\Teff$ and $0.005$ in $\log(R/R_\odot)$.  \label{figure:nn_validation}}
\end{figure}

We use the code \misty\footnote{https://github.com/pacargile/MISTy} (Cargile 2026, in prep.) to train a neural network emulator on our model grids to allow for very quick evaluation and likelihood optimization.  They are trained on the input labels $(\mathrm{EEP}, ~\FeHi, ~\Mi, ~\amlt)$ and the output labels of $(\log(\Teff),~\log(L),~\log(t),~\Delta M,~\Delta\FeHs)$ where $t$ is the age, $\Delta M$ is the accumulated mass loss, and $\Delta\FeHs$ is the change in surface metallicity $\FeHs$ relative to initial metallicity $\FeHi$.  MIST accounts for diffusion of all elemental species and therefore the surface metallicity can be depleted relative to the initial bulk metallicity \citep{Choi_2016, Dotter_2017, Dotter_2026}.  Transforming from the $\Delta$ quantities back to the observables $M,~\FeHs$ is trivial using the input labels.  We found that these transformations aided in training, allowing us to use a less complex network to emulate the behavior of stellar evolution.  The specific network architecture used for all three of the grids is a Multi-Layer Perceptron with three hidden layers of 64 neurons each, using a Huber Loss function.  Figure \ref{figure:nn_validation} compares the NN emulation to the actual stellar tracks of the metal-rich grid, demonstrating that the emulation is accurate and safe to use during the fitting procedure.  

\subsection{Isochrone Fitting Procedure}
\label{subsection:fitting}
For each system we define the likelihood $\mathcal{L}(\mathbf{D}|\boldsymbol\theta)$ over the set of available observables $\mathbf{D}$ and the input model parameters $\boldsymbol\theta$.  Our data are a mix of single stars, binary stars, and clusters containing multiple binaries.  We assume all of the components within a multiple system are co-natal, that is they were born at the same time with the same initial composition.  The input parameters $\boldsymbol\theta$ are then the system's common age $t$, common initial metallicity $\FeHi$, and an initial mass $\Mi$ and mixing length $\amlt$ for each component.  

The observables $\mathbf{D}$ vary depending on the type and multiplicity of the particular system.  The observables $\mathbf{D}$ for systems in the \debs\ sample, including those in clusters, are the dynamical mass $M$, radius $R$, temperature $\Teff$, and surface metallicity $\FeHs$ for each component.  The \interferometric\ stars do not have dynamical mass measurements so we instead apply a Kroupa IMF prior \citep{Kroupa_2001} during their fits.  The majority of the \interferometric\ sample are single stars, but the systems 16 Cyg and 61 Cyg are two unique cases being visual binaries where both the A \& B components have measured radii.  These systems' orbits are too long to provide useful dynamical mass constraints \citep{Hauser_1999, Kervella_2021} but we do leverage their binarity and require both components to share an isochrone as we do for the \debs.  Thus the observables for the \interferometric\ systems are the radius $R$, temperature $\Teff$, and surface metallicity $\FeHs$ for each component.

The likelihood for each star is formed by attributing a Gaussian term to each of the observables, using the uncertainties reported in DEBCat and GBSv3.  The likelihood of a single star is then the product of all of its observables' terms, and the likelihood for the entire system is the product of all member stars' likelihoods.  An example likelihood for an eclipsing binary is given in Equation \ref{equation:likelihood_term}, with $\tilde D(\boldsymbol\theta)$ representing the model prediction for a given observable as a function of the input parameters $\boldsymbol\theta$.  The model predictions are made using the NN emulator appropriate for the system's observed metallicity.  In the regions of overlap the more metal-rich network is generally adopted (see Figure \ref{figure:data_and_model_grid_coverage}), except for $\mu$ Cas and the globular cluster 47 Tuc which instead adopt the very metal-poor network. The choice of NN in the overlap regions amounts to a choice of $\aFei$, and in Section \ref{section:caveats} we explore the impact of this choice on the inference of $\amlt$.

\begin{align}
\label{equation:likelihood_term}
\mathcal{L}_\mathrm{DEB}(R_A|\boldsymbol\theta) &\propto \exp\left\{-\frac{1}{2}\left(\frac{R_A-\tilde R_A(\boldsymbol\theta)}{\sigma_{R_A}}\right)^2\right\} \nonumber\\
\mathcal{L}_\mathrm{DEB}(\mathbf{D_A}|\boldsymbol\theta) &= \mathcal{L}_{M_A}\mathcal{L}_{R_A}\mathcal{L}_{T_A}\mathcal{L}_{{\FeH}_A} \nonumber\\
\mathcal{L}_\mathrm{DEB}(\mathbf{D}|\boldsymbol\theta) &= \mathcal{L}_A\mathcal{L}_B \nonumber \\
\end{align}

Using the system age directly as an input parameter is often unstable during short lived post-main sequence phases.  We therefore replace the system age with the EEP of the system's largest, i.e., most evolved component when we perform our inferences.  The mapping between age and EEP is conditional on a star's other parameters ($t = t(\mathrm{EEP}|\Mi,~\FeHi,~\amlt)$) but it is always one$-$to$-$one, so enforcing the other members of a binary or cluster to share the same age as this `anchor' star can be done by composing the mappings $\mathrm{EEP}_\mathrm{anchor}\rightarrow t_\mathrm{system}\rightarrow \mathrm{EEP}_\mathrm{companion}$.  This $\mathrm{EEP}_\mathrm{companion}$ then enters into $\boldsymbol\theta$ when calculating $\tilde D_\mathrm{companion}$ in the likelihood.
Switching to EEP also requires the inclusion of a prior $\propto \frac{d\mathrm{Age}}{d\mathrm{EEP}}$ so that samples are drawn uniformly in age rather than uniformly in EEP \citep[see][]{Cargile_2020}.

We place no further priors on a system's age, in particular we do not enforce the inferred age to be younger than the age of the universe \citep[13.8 Gyr;][]{PlanckCollaboration_2020}.  Isochrone fits that infer unphysically old ages are signs of missing model physics.  All other input parameters are given a uniform prior over the valid domain of the NN emulator.  

With the likelihood and prior functions defined we sample each system's posterior using \texttt{emcee}\footnote{https://emcee.readthedocs.io} \citep{Foreman-Mackey_2013}.  Model predictions $\tilde D(\boldsymbol\theta)$ are made using the NN emulator appropriate for the system's metallicity.  We run two versions of every system's inference:  one version that enforces $\amlt = \solaramlt$ and another that allows $\amlt$ to vary as a free parameter over the entire range $[1.0,~3.0]$.  Doing two runs in this way allows us to confirm which systems are inconsistent with the $\amlt = \solaramlt$ assumption and then explore the role of a non-solar mixing length in explaining this inconsistency.  All sample chains and posteriors have been manually inspected to ensure they are converged and well behaved.

\section{Mixing Length in Detached Eclipsing Binaries}
\label{section:debs_results}

\subsection{Failures of $\solaramlt$ Models}
\label{subsection:solaramlt_failures}
Table \ref{table:debs_results} lists the results of isochrone fits to our \debs\ sample, using either models with the standard $\amlt=\solaramlt$ assumption or with models that allow $\amlt$ to vary.  Columns 2 and 3 list the respective $\chi^2$ for the maximum a posteriori (MAP) sample in our MCMC chains (denoted as $\hat{\boldsymbol\theta}$).  The $\solaramlt$ fit has model parameters $\boldsymbol\theta = \{\mathrm{EEP}_0, ~\FeHi, ~M_{i,A}, ~M_{i,B} \}$ and thus has $\nu = 4$ degrees of freedom, while the variable $\amlt$ run also includes $\{ \amltA,~\amltB \}$ in $\boldsymbol\theta$ and thus has $\nu = 2$.  The fits to Procyon and $\mu$ Cas only consider the primary star and so $\nu = 1$ and $\nu = 0$ for the $\solaramlt$ and variable $\amlt$ fits respectively.  The $\chi^2$ of the $\solaramlt$ and variable $\amlt$ fits cannot be directly compared because of these additional free parameters, so we also report the difference in Bayesian Information Criterion ($\Delta\mathrm{BIC} = \mathrm{BIC}_{\solaramlt} - \mathrm{BIC}_{\amlt}$) between the two.  Both runs share the same eight observables in the likelihood and thus a $\Delta\mathrm{BIC} \gtrsim 6$ indicates a strong preference for the models with a variable $\amlt$.  

\begin{figure}[t!]
\epsscale{1.0}
\plotone{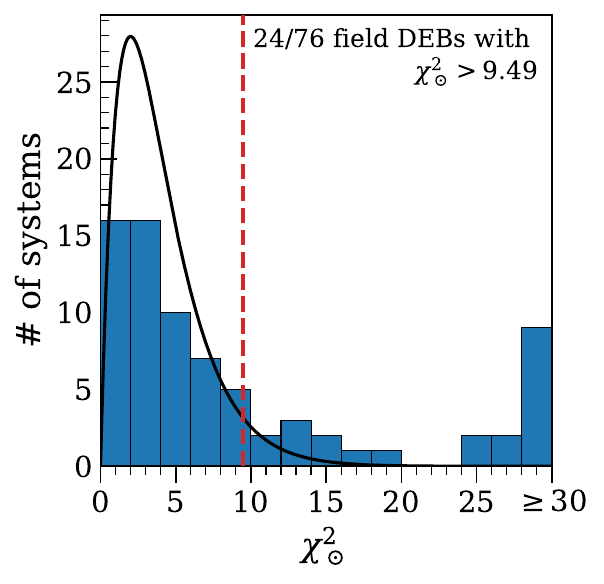}
\caption{$\chi^2$ of the \debs\ isochrone fits using standard $\solaramlt$ models.  The black line is the expected $\chi^2_{\nu}$ distribution for models with $\nu = 4$ degrees of freedom.  If $\solaramlt$ models could faithfully reproduce the observations of the \debs\ then we would expect the histogram to approximately follow the black line.  The clear excess of systems with $\chi^2_\odot > 9.49$ strongly indicates some flaw or additional physics not captured in the $\solaramlt$ models. \label{figure:debs_solaramlt_chi2}}
\end{figure}

Figure \ref{figure:debs_solaramlt_chi2} plots the distribution of $\chi^2$ for the $\solaramlt$ runs.  We find that $\sim1/3$ of the \debs\ systems are poorly fit by models assuming $\solaramlt$, a fraction far larger than expected by random chance.  In an attempt to classify the $\solaramlt$ failures we flag systems whose $\solaramlt$ fits fail one way or another at a ``$2\sigma$" or 95\% confidence level.  We would like to identify interesting candidates for future case studies so we intentionally choose a threshold that is not too strict, but we must therefore be cautious when reading into any one flagged system and arguing for an issue in the models.  An underestimation of a system's uncertainty or an otherwise biased measurement could just as easily drive a large $\chi^2$.  Still, it is unlikely that an entire third of the \debs\ have problematic data and so many of these cases are bound to be genuine signals of model imperfections.

We consider four failure modes in the $\solaramlt$ fits:  a system $\chi^2_\odot > 9.49$, a temperature residual $|\Teff - \tilde T_\mathrm{eff} (\hat{\boldsymbol\theta})| > 2\sigma_T$ on either component in the system, a radius residual $|R - \tilde R (\hat{\boldsymbol\theta})| > 2\sigma_R$ on either component in the system, or if 95\% of the system's marginal age posterior is older than the age of the universe \citep[$t_5~> 13.8$ Gyr;][]{PlanckCollaboration_2020}.  A $\chi^2_\odot > 9.49$ exceeds the 95th percentile of a $\chi^2$ distribution with $\nu = 4$ degrees of freedom, appropriate for the $\solaramlt$ fits.  The last failure mode is useful to catch instances of radius inflation where all realistically aged models are too small to match the observed radius.  The $\tilde T, ~\tilde R$, and $~t_5$ flags are recorded in Column 6 of Table \ref{table:debs_results}, with the $\Teff$ and $R$ residual flags indicating the direction of mismatch between model and observations e.g., $\tilde T\downarrow$ indicates that the best fitting $\solaramlt$ model predicts a $\Teff$ cooler than the observed temperature.

\begin{figure*}[t!]
\epsscale{1.0}
\plotone{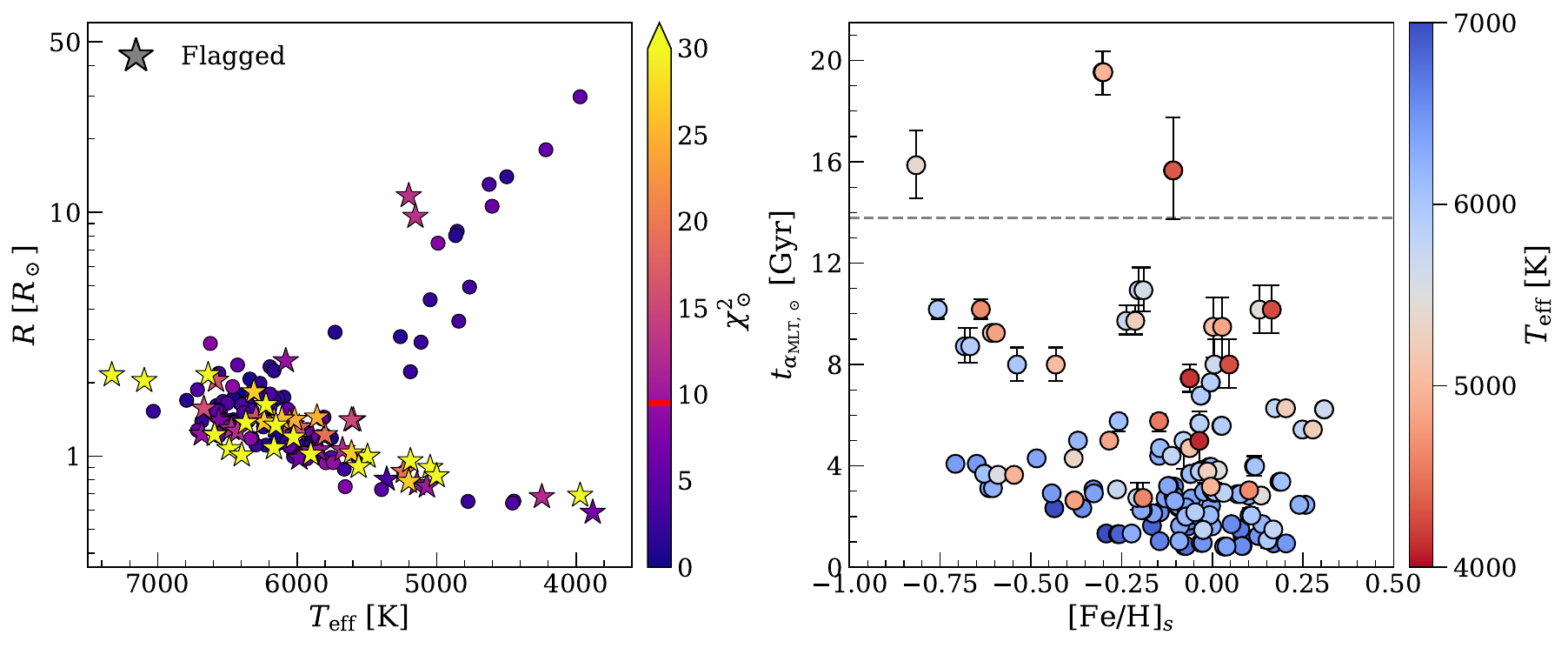}
\caption{\debs\ sample plotted in $\Teff-\log(R)$ and $\FeHs-t$ space.  On the left stars are colored by the $\chi^2$ of their system's $\solaramlt$ fit found in Table \ref{table:debs_results}.  Flagged systems ($\chi^2_\odot > 9.49$) are marked as stars.  Three members of the DEBs sample have unphysical ages, indicating some flaw with the age scale of our models.  \label{figure:debs_solaramlt_kiel}}
\end{figure*}

Figure \ref{figure:debs_solaramlt_kiel} plots the results of the \debs\ $\solaramlt$ fits in the HR diagram and $\FeHs - t$ space.  The flagged systems are largely restricted to the main sequence, with an increased likelihood for cool stars ($\Teff \lesssim 5400$ K) to be flagged.  Several of these cool stars tend towards unusually old ages despite being metal-rich solar neighborhood stars, but their scatter is large enough that we cannot claim a general issue with $\solaramlt$ models at low $\Teff$ on the basis of ages.  

There are however three particular systems that have fitted ages $> 13.8$ Gyr.  From low to high metallicity they are $\mu$ Cas, TIC 17303250, and TIC 97729372, with only TIC 17303250 fully satisfying the age flag with $t_5~> 13.8$ Gyr.  TIC 17303250 is a slightly metal-poor system of two $\sim 0.7 M_\odot$ stars that can only be fit well by $\solaramlt$ models at unphysically old ages.  If we force $t~< 13.8$ Gyr then the $\solaramlt$ fit becomes dramatically worse, with  $\chi^2 = 62$.  On the assumption that TIC 17303250's measurements are sound then it is a clear indiction of radius inflation.  TIC 97729372 is a similar but slightly less severe case as a pair of $0.64 M_\odot$ dwarfs, and $\mu$ Cas A is a metal-poor dwarf $(\sim 0.74 M_\odot)$ that very nearly satisfies the age flag, again suggesting some form of radius inflation.

The ``$\amlt/\solaramlt$" and ``95\% CI" columns in Table \ref{table:debs_results} summarize the results of the variable $\amlt$ isochrone fits, reporting the median and 95\% credible interval of the inferred mixing length in each component.  We report only a lower (upper) bound if the $\amlt$ posterior is slammed up against a prior edge and the 95\% credible interval is entirely below (above) $\amlt/\solaramlt = 1$.  If the width of the 95\% credible interval exceeds 66.7\% of the total prior range then we consider the $\amlt$ inference to be uninformative and do not report anything.  

The last column, ``$\alpha_A$ vs. $\alpha_B$", considers the joint mixing length posterior between the two components.  Here we flag systems where the 95\% interval of the difference $\amltA - \amltB$ is either entirely above or below $\amltA - \amltB = 0$, indicating that $\amltA \neq \amltB$ at high confidence.  This is useful because there are systems where both components' marginal $\amlt$ posteriors appear consistent with $\solaramlt$ when considered individually, but the covariance between $\amltA$ and $\amltB$ ensures that both cannot equal $\solaramlt$ \emph{simultaneously}.  A good example of this can be seen in $\alpha$ Centauri under the \citet{Kervella_2016, Kervella_2017} measurements (Section \ref{subsection:discuss_alphaCen}).

\startlongtable
\begin{deluxetable*}{lCCC|cCCCC}
\tablecaption{Summary of the \debs\ Isochrone Fits\label{table:debs_results}}
\tablehead{
\colhead{System} & \colhead{$\chi^2_\odot$} & \colhead{$\chi^2_{\amlt}$} & \colhead{$\Delta$BIC} &
\colhead{Comp.} & \colhead{$\solaramlt$ Failure mode} & \colhead{$\amlt/\solaramlt$} & \colhead{$95\%$ CI} & \colhead{$\amltA$ vs.\ $\amltB$}
}
\startdata
V636 Cen & 112.4 & 0.7 & 107.6 & A & \phantom{T\uparrow}~~\tilde{R}\uparrow~~\phantom{t_5} & 0.83 & [0.67,\,1.03] & \amltA > \amltB \\
 &  &  &  & B$^I$ & \tilde{T}\uparrow~~\tilde{R}\downarrow~~\phantom{t_5} & < 0.60 &  &  \\
V375 Cep & 112.2 & 1.9 & 106.2 & A & \tilde{T}\downarrow~~\phantom{R\uparrow}~~\phantom{t_5} & 1.27 & [0.98,\,1.46] & \amltA > \amltB \\
 &  &  &  & B$^I$ & \tilde{T}\uparrow~~\tilde{R}\downarrow~~\phantom{t_5} & < 0.56 &  &  \\
KIC 9850387 & 85.0 & 9.7 & 71.2 & A & \phantom{T\uparrow}~~\phantom{R\uparrow}~~\phantom{t_5} & \nodata & \nodata & \amltA > \amltB \\
 &  &  &  & B & \phantom{T\uparrow}~~\tilde{R}\downarrow~~\phantom{t_5} & 0.74 & [0.69,\,0.80] &  \\
EF Aqr & 55.5 & 0.5 & 50.8 & A & \phantom{T\uparrow}~~\phantom{R\uparrow}~~\phantom{t_5} & 0.74 & [0.49,\,1.08] & \amltA > \amltB \\
 &  &  &  & B$^I$ & \tilde{T}\uparrow~~\tilde{R}\downarrow~~\phantom{t_5} & < 0.61 &  &  \\
AL Ari & 47.2 & 1.6 & 41.4 & A & \tilde{T}\uparrow~~\phantom{R\uparrow}~~\phantom{t_5} & < 0.85 &  &  \\
 &  &  &  & B & \tilde{T}\uparrow~~\tilde{R}\downarrow~~\phantom{t_5} & 0.66 & [0.59,\,0.73] &  \\
HD 149946 & 42.9 & 0.8 & 37.9 & A & \tilde{T}\uparrow~~\phantom{R\uparrow}~~\phantom{t_5} & \nodata & \nodata &  \\
 &  &  &  & B & \phantom{T\uparrow}~~\phantom{R\uparrow}~~\phantom{t_5} & 0.98 & [0.88,\,1.10] &  \\
TYC 5962-2159-1 & 35.9 & 9.1 & 22.6 & A & \phantom{T\uparrow}~~\phantom{R\uparrow}~~\phantom{t_5} & \nodata & \nodata & \amltA > \amltB \\
 &  &  &  & B$^I$ & \tilde{T}\uparrow~~\tilde{R}\downarrow~~\phantom{t_5} & < 0.60 &  &  \\
KIC 6525196 & 32.8 & 8.0 & 20.6 & A & \phantom{T\uparrow}~~\phantom{R\uparrow}~~\phantom{t_5} & > 1.12 &  & \amltA > \amltB \\
 &  &  &  & B & \phantom{T\uparrow}~~\phantom{R\uparrow}~~\phantom{t_5} & 0.92 & [0.76,\,1.08] &  \\
HIP 7666 & 31.6 & 10.2 & 17.2 & A & \tilde{T}\downarrow~~\phantom{R\uparrow}~~\phantom{t_5} & \nodata & \nodata &  \\
 &  &  &  & B & \tilde{T}\uparrow~~\phantom{R\uparrow}~~\phantom{t_5} & 0.73 & [0.63,\,0.83] &  \\
Kepler-35 & 26.9 & 4.9 & 17.8 & A & \phantom{T\uparrow}~~\phantom{R\uparrow}~~\phantom{t_5} & 0.67 & [0.49,\,0.88] & \amltA < \amltB \\
 &  &  &  & B & \phantom{T\uparrow}~~\phantom{R\uparrow}~~\phantom{t_5} & 1.18 & [0.92,\,1.48] &  \\
BN Scl & 26.8 & 20.1 & 2.5 & A & \tilde{T}\uparrow~~\phantom{R\uparrow}~~\phantom{t_5} & < 0.92 &  & \amltA < \amltB \\
 &  &  &  & B & \phantom{T\uparrow}~~\phantom{R\uparrow}~~\phantom{t_5} & 0.82 & [0.73,\,0.95] &  \\
WZ Oph & 25.5 & 7.7 & 13.7 & A & \phantom{T\uparrow}~~\phantom{R\uparrow}~~\phantom{t_5} & < 0.66 &  &  \\
 &  &  &  & B & \tilde{T}\uparrow~~\phantom{R\uparrow}~~\phantom{t_5} & 0.60 & [0.49,\,0.73] &  \\
V785 Cep & 24.6 & 8.1 & 12.3 & A & \tilde{T}\uparrow~~\phantom{R\uparrow}~~\phantom{t_5} & 0.85 & [0.70,\,1.04] & \amltA < \amltB \\
 &  &  &  & B & \tilde{T}\downarrow~~\phantom{R\uparrow}~~\phantom{t_5} & 1.03 & [0.85,\,1.26] &  \\
$\alpha$ Cen$^{*}$ & 19.1 & 11.8 & 3.2 & A & \phantom{T\uparrow}~~\phantom{R\uparrow}~~\phantom{t_5} & 0.93 & [0.83,\,1.05] & \amltA < \amltB \\
 &  &  &  & B & \phantom{T\uparrow}~~\tilde{R}\uparrow~~\phantom{t_5} & 1.04 & [0.95,\,1.14] &  \\
TIC 172900988 & 17.7 & 0.1 & 13.5 & A & \phantom{T\uparrow}~~\tilde{R}\uparrow~~\phantom{t_5} & \nodata & \nodata & \amltA > \amltB \\
 &  &  &  & B & \phantom{T\uparrow}~~\phantom{R\uparrow}~~\phantom{t_5} & 1.06 & [0.70,\,1.33] &  \\
\shortstack[l]{Gaia DR3 \\ 5630315754100784000} & 15.7 & 5.9 & 5.6 & A & \phantom{T\uparrow}~~\phantom{R\uparrow}~~\phantom{t_5} & 1.03 & [0.54,\,1.47] &  \\
 &  &  &  & B & \tilde{T}\uparrow~~\tilde{R}\downarrow~~\phantom{t_5} & 0.82 & [0.68,\,1.05] &  \\
NY Hya & 14.9 & 0.5 & 10.2 & A & \phantom{T\uparrow}~~\phantom{R\uparrow}~~\phantom{t_5} & < 0.71 &  &  \\
 &  &  &  & B & \phantom{T\uparrow}~~\phantom{R\uparrow}~~\phantom{t_5} & < 0.69 &  &  \\
TIC 81462274 & 12.8 & 1.5 & 7.1 & A & \phantom{T\uparrow}~~\phantom{R\uparrow}~~\phantom{t_5} & > 0.94 &  &  \\
 &  &  &  & B & \phantom{T\uparrow}~~\phantom{R\uparrow}~~\phantom{t_5} & 1.08 & [0.87,\,1.47] &  \\
V505 Per & 12.4 & 2.6 & 5.6 & A & \phantom{T\uparrow}~~\phantom{R\uparrow}~~\phantom{t_5} & 0.73 & [0.62,\,0.88] & \amltA < \amltB \\
 &  &  &  & B & \phantom{T\uparrow}~~\phantom{R\uparrow}~~\phantom{t_5} & 0.78 & [0.70,\,0.91] &  \\
IM Vir & 12.2 & 3.6 & 4.4 & A & \phantom{T\uparrow}~~\phantom{R\uparrow}~~\phantom{t_5} & 0.97 & [0.63,\,1.42] &  \\
 &  &  &  & B$^I$ & \phantom{T\uparrow}~~\tilde{R}\downarrow~~\phantom{t_5} & < 1.01 &  &  \\
UW LMi & 10.4 & 0.1 & 6.1 & A & \phantom{T\uparrow}~~\phantom{R\uparrow}~~\phantom{t_5} & 0.67 & [0.51,\,0.86] &  \\
 &  &  &  & B & \phantom{T\uparrow}~~\phantom{R\uparrow}~~\phantom{t_5} & 0.71 & [0.55,\,0.88] &  \\
TIC 17303250 & 10.1 & 17.5 & -11.6 & A$^I$ & \phantom{T\uparrow}~~\phantom{R\uparrow}~~t_5 & 0.66 & [0.56,\,0.78] &  \\
 &  &  &  & B$^I$ & \phantom{T\uparrow}~~\phantom{R\uparrow}~~t_5 & < 0.65 &  &  \\
KX Cnc & 9.9 & 19.1 & -13.4 & A & \phantom{T\uparrow}~~\phantom{R\uparrow}~~\phantom{t_5} & 0.88 & [0.76,\,1.05] & \amltA < \amltB \\
 &  &  &  & B & \phantom{T\uparrow}~~\phantom{R\uparrow}~~\phantom{t_5} & 0.99 & [0.84,\,1.20] &  \\
V432 Aur & 9.5 & 1.3 & 4.1 & A & \tilde{T}\downarrow~~\phantom{R\uparrow}~~\phantom{t_5} & > 1.15 &  &  \\
 &  &  &  & B & \phantom{T\uparrow}~~\phantom{R\uparrow}~~\phantom{t_5} & \nodata & \nodata &  \\
KIC 7177553 & 9.1 & 24.6 & -19.7 & A & \phantom{T\uparrow}~~\phantom{R\uparrow}~~\phantom{t_5} & 0.84 & [0.62,\,1.09] & \amltA < \amltB \\
 &  &  &  & B & \phantom{T\uparrow}~~\phantom{R\uparrow}~~\phantom{t_5} & 1.15 & [0.90,\,1.47] &  \\
CPD -54 810 & 8.7 & 4.1 & 0.4 & A & \phantom{T\uparrow}~~\phantom{R\uparrow}~~\phantom{t_5} & 1.05 & [0.59,\,1.15] &  \\
 &  &  &  & B & \phantom{T\uparrow}~~\phantom{R\uparrow}~~\phantom{t_5} & 0.98 & [0.85,\,1.14] &  \\
BD +37 410 & 8.2 & 7.3 & -3.3 & A & \phantom{T\uparrow}~~\phantom{R\uparrow}~~\phantom{t_5} & \nodata & \nodata &  \\
 &  &  &  & B & \phantom{T\uparrow}~~\phantom{R\uparrow}~~\phantom{t_5} & 0.95 & [0.86,\,1.05] &  \\
KIC 8430105 & 7.9 & 1.9 & 1.9 & A & \phantom{T\uparrow}~~\phantom{R\uparrow}~~\phantom{t_5} & 1.04 & [0.88,\,1.23] &  \\
 &  &  &  & B & \phantom{T\uparrow}~~\phantom{R\uparrow}~~\phantom{t_5} & 1.28 & [1.09,\,1.48] &  \\
LX Mus & 7.9 & 0.7 & 3.0 & A & \phantom{T\uparrow}~~\tilde{R}\uparrow~~\phantom{t_5} & 1.07 & [0.80,\,1.39] & \amltA > \amltB \\
 &  &  &  & B & \phantom{T\uparrow}~~\phantom{R\uparrow}~~\phantom{t_5} & 0.92 & [0.64,\,1.22] &  \\
V530 Ori & 6.9 & 3.0 & -0.3 & A & \phantom{T\uparrow}~~\phantom{R\uparrow}~~\phantom{t_5} & > 1.12 &  & \amltA > \amltB \\
 &  &  &  & B$^I$ & \tilde{T}\uparrow~~\phantom{R\uparrow}~~\phantom{t_5} & 0.84 & [0.70,\,1.13] &  \\
TYC 5227-1023-1 & 6.7 & 2.1 & 0.4 & A & \phantom{T\uparrow}~~\phantom{R\uparrow}~~\phantom{t_5} & \nodata & \nodata &  \\
 &  &  &  & B & \phantom{T\uparrow}~~\phantom{R\uparrow}~~\phantom{t_5} & \nodata & \nodata &  \\
QR Hya & 6.5 & 0.1 & 2.1 & A & \phantom{T\uparrow}~~\phantom{R\uparrow}~~\phantom{t_5} & 0.89 & [0.68,\,1.13] &  \\
 &  &  &  & B & \phantom{T\uparrow}~~\phantom{R\uparrow}~~\phantom{t_5} & 0.84 & [0.68,\,1.02] &  \\
KIC 8410637 & 6.2 & 4.0 & -2.0 & A & \phantom{T\uparrow}~~\phantom{R\uparrow}~~\phantom{t_5} & 0.93 & [0.81,\,1.04] & \amltA < \amltB \\
 &  &  &  & B & \phantom{T\uparrow}~~\phantom{R\uparrow}~~\phantom{t_5} & 1.23 & [0.96,\,1.48] &  \\
V963 Cen & 6.1 & 0.3 & 1.6 & A & \phantom{T\uparrow}~~\phantom{R\uparrow}~~\phantom{t_5} & 0.80 & [0.65,\,0.98] &  \\
 &  &  &  & B & \phantom{T\uparrow}~~\phantom{R\uparrow}~~\phantom{t_5} & 0.79 & [0.65,\,0.95] &  \\
KIC 7821010 & 5.9 & 1.5 & 0.3 & A & \phantom{T\uparrow}~~\phantom{R\uparrow}~~\phantom{t_5} & > 1.02 &  &  \\
 &  &  &  & B & \phantom{T\uparrow}~~\phantom{R\uparrow}~~\phantom{t_5} & 1.20 & [0.94,\,1.48] &  \\
AL Dor & 5.6 & 0.1 & 1.4 & A & \phantom{T\uparrow}~~\phantom{R\uparrow}~~\phantom{t_5} & 0.87 & [0.71,\,1.04] &  \\
 &  &  &  & B & \phantom{T\uparrow}~~\phantom{R\uparrow}~~\phantom{t_5} & 0.84 & [0.69,\,1.01] &  \\
\shortstack[l]{Gaia DR3 \\ 4063342909579750528} & 5.6 & 4.4 & -2.9 & A & \phantom{T\uparrow}~~\phantom{R\uparrow}~~\phantom{t_5} & 0.81 & [0.64,\,1.02] &  \\
 &  &  &  & B & \phantom{T\uparrow}~~\phantom{R\uparrow}~~\phantom{t_5} & 0.66 & [0.57,\,0.74] &  \\
V338 Vir & 5.3 & 3.5 & -2.3 & A & \phantom{T\uparrow}~~\phantom{R\uparrow}~~\phantom{t_5} & 0.97 & [0.83,\,1.16] &  \\
 &  &  &  & B & \phantom{T\uparrow}~~\phantom{R\uparrow}~~\phantom{t_5} & 0.94 & [0.51,\,1.13] &  \\
TYC 7091-888-1 & 4.8 & 2.3 & -1.7 & A & \phantom{T\uparrow}~~\phantom{R\uparrow}~~\phantom{t_5} & 1.12 & [0.81,\,1.46] &  \\
 &  &  &  & B & \phantom{T\uparrow}~~\phantom{R\uparrow}~~\phantom{t_5} & 1.22 & [1.00,\,1.48] &  \\
$\alpha$ Cen$^\dagger$ & 4.6 & 0.9 & -0.4 & A & \phantom{T\uparrow}~~\phantom{R\uparrow}~~\phantom{t_5} & 1.07 & [0.96,\,1.18] &  \\
 &  &  &  & B & \phantom{T\uparrow}~~\phantom{R\uparrow}~~\phantom{t_5} & 1.09 & [1.00,\,1.17] &  \\
KIC 3439031 & 4.5 & 9.8 & -9.4 & A & \phantom{T\uparrow}~~\phantom{R\uparrow}~~\phantom{t_5} & \nodata & \nodata &  \\
 &  &  &  & B & \phantom{T\uparrow}~~\phantom{R\uparrow}~~\phantom{t_5} & 1.24 & [0.90,\,1.48] &  \\
V570 Per & 4.5 & 1.0 & -0.7 & A & \phantom{T\uparrow}~~\phantom{R\uparrow}~~\phantom{t_5} & 0.74 & [0.49,\,1.04] & \amltA < \amltB \\
 &  &  &  & B & \phantom{T\uparrow}~~\phantom{R\uparrow}~~\phantom{t_5} & 0.94 & [0.81,\,1.12] &  \\
TIC 97729372 & 4.2 & 2.9 & -2.9 & A & \phantom{T\uparrow}~~\phantom{R\uparrow}~~\phantom{t_5} & \nodata & \nodata &  \\
 &  &  &  & B & \phantom{T\uparrow}~~\phantom{R\uparrow}~~\phantom{t_5} & \nodata & \nodata &  \\
HD 32129 & 4.1 & 1.4 & -1.5 & A & \phantom{T\uparrow}~~\phantom{R\uparrow}~~\phantom{t_5} & \nodata & \nodata &  \\
 &  &  &  & B & \phantom{T\uparrow}~~\phantom{R\uparrow}~~\phantom{t_5} & 0.97 & [0.89,\,1.06] &  \\
ASAS J171750-1915.3 & 3.9 & 0.6 & -0.9 & A & \phantom{T\uparrow}~~\phantom{R\uparrow}~~\phantom{t_5} & \nodata & \nodata &  \\
 &  &  &  & B & \phantom{T\uparrow}~~\phantom{R\uparrow}~~\phantom{t_5} & \nodata & \nodata &  \\
CD Tau & 3.6 & 0.6 & -1.2 & A & \phantom{T\uparrow}~~\phantom{R\uparrow}~~\phantom{t_5} & \nodata & \nodata &  \\
 &  &  &  & B & \phantom{T\uparrow}~~\phantom{R\uparrow}~~\phantom{t_5} & 0.82 & [0.61,\,1.21] &  \\
FM Leo & 3.5 & 0.3 & -1.0 & A & \phantom{T\uparrow}~~\phantom{R\uparrow}~~\phantom{t_5} & 0.98 & [0.80,\,1.43] &  \\
 &  &  &  & B & \phantom{T\uparrow}~~\phantom{R\uparrow}~~\phantom{t_5} & 0.96 & [0.73,\,1.21] &  \\
KIC 10001167 & 3.3 & 1.4 & -2.2 & A & \phantom{T\uparrow}~~\phantom{R\uparrow}~~\phantom{t_5} & 0.95 & [0.87,\,1.03] &  \\
 &  &  &  & B & \phantom{T\uparrow}~~\phantom{R\uparrow}~~\phantom{t_5} & 1.01 & [0.76,\,1.33] &  \\
KIC 11235323 & 3.3 & 1.7 & -2.6 & A & \phantom{T\uparrow}~~\phantom{R\uparrow}~~\phantom{t_5} & 0.82 & [0.50,\,1.21] &  \\
 &  &  &  & B & \phantom{T\uparrow}~~\phantom{R\uparrow}~~\phantom{t_5} & \nodata & \nodata &  \\
RU Cnc & 2.8 & 1.2 & -2.6 & A & \phantom{T\uparrow}~~\phantom{R\uparrow}~~\phantom{t_5} & \nodata & \nodata &  \\
 &  &  &  & B$^I$ & \phantom{T\uparrow}~~\phantom{R\uparrow}~~\phantom{t_5} & 0.72 & [0.55,\,0.94] &  \\
LL Aqr & 2.6 & 0.3 & -1.8 & A & \phantom{T\uparrow}~~\phantom{R\uparrow}~~\phantom{t_5} & 1.06 & [0.84,\,1.38] &  \\
 &  &  &  & B & \phantom{T\uparrow}~~\phantom{R\uparrow}~~\phantom{t_5} & 1.05 & [0.92,\,1.21] &  \\
AI Phe & 2.6 & 0.7 & -2.3 & A & \phantom{T\uparrow}~~\phantom{R\uparrow}~~\phantom{t_5} & 1.21 & [1.03,\,1.40] & \amltA > \amltB \\
 &  &  &  & B & \phantom{T\uparrow}~~\phantom{R\uparrow}~~\phantom{t_5} & 1.01 & [0.93,\,1.10] &  \\
VZ Hya & 2.5 & 0.7 & -2.3 & A & \phantom{T\uparrow}~~\phantom{R\uparrow}~~\phantom{t_5} & \nodata & \nodata &  \\
 &  &  &  & B & \phantom{T\uparrow}~~\phantom{R\uparrow}~~\phantom{t_5} & 0.97 & [0.77,\,1.24] &  \\
KIC 6131659 & 2.5 & 0.1 & -1.8 & A & \phantom{T\uparrow}~~\phantom{R\uparrow}~~\phantom{t_5} & 0.99 & [0.66,\,1.35] &  \\
 &  &  &  & B & \phantom{T\uparrow}~~\phantom{R\uparrow}~~\phantom{t_5} & \nodata & \nodata &  \\
EW Ori & 2.5 & 0.1 & -1.8 & A & \phantom{T\uparrow}~~\phantom{R\uparrow}~~\phantom{t_5} & 0.87 & [0.66,\,1.17] &  \\
 &  &  &  & B & \phantom{T\uparrow}~~\phantom{R\uparrow}~~\phantom{t_5} & 0.85 & [0.67,\,1.07] &  \\
CO And & 2.4 & 0.8 & -2.6 & A & \phantom{T\uparrow}~~\phantom{R\uparrow}~~\phantom{t_5} & 0.99 & [0.61,\,1.29] &  \\
 &  &  &  & B & \phantom{T\uparrow}~~\phantom{R\uparrow}~~\phantom{t_5} & 0.86 & [0.57,\,1.10] &  \\
ASAS J090232-5653.4 & 2.4 & 1.3 & -3.1 & A & \phantom{T\uparrow}~~\phantom{R\uparrow}~~\phantom{t_5} & \nodata & \nodata &  \\
 &  &  &  & B & \phantom{T\uparrow}~~\phantom{R\uparrow}~~\phantom{t_5} & \nodata & \nodata &  \\
Kepler-1647 & 2.3 & 1.3 & -3.1 & A & \phantom{T\uparrow}~~\phantom{R\uparrow}~~\phantom{t_5} & 1.21 & [0.98,\,1.47] & \amltA > \amltB \\
 &  &  &  & B & \phantom{T\uparrow}~~\phantom{R\uparrow}~~\phantom{t_5} & 0.96 & [0.85,\,1.09] &  \\
\shortstack[l]{Gaia DR3 \\ 26100417474249344} & 2.3 & 1.2 & -3.0 & A & \phantom{T\uparrow}~~\phantom{R\uparrow}~~\phantom{t_5} & 1.24 & [0.94,\,1.48] &  \\
 &  &  &  & B & \phantom{T\uparrow}~~\phantom{R\uparrow}~~\phantom{t_5} & 1.15 & [0.85,\,1.44] &  \\
TIC 71877648 & 2.2 & 0.4 & -2.3 & A & \phantom{T\uparrow}~~\phantom{R\uparrow}~~\phantom{t_5} & \nodata & \nodata &  \\
 &  &  &  & B & \phantom{T\uparrow}~~\phantom{R\uparrow}~~\phantom{t_5} & \nodata & \nodata &  \\
BK Peg & 1.7 & 0.8 & -3.3 & A & \phantom{T\uparrow}~~\phantom{R\uparrow}~~\phantom{t_5} & 0.93 & [0.52,\,1.21] &  \\
 &  &  &  & B & \phantom{T\uparrow}~~\phantom{R\uparrow}~~\phantom{t_5} & 0.86 & [0.68,\,1.06] &  \\
KIC 7037405 & 1.7 & 1.0 & -3.5 & A & \phantom{T\uparrow}~~\phantom{R\uparrow}~~\phantom{t_5} & 0.91 & [0.77,\,1.07] &  \\
 &  &  &  & B & \phantom{T\uparrow}~~\phantom{R\uparrow}~~\phantom{t_5} & 1.08 & [0.63,\,1.43] &  \\
GX Gem & 1.6 & 0.3 & -2.8 & A & \phantom{T\uparrow}~~\phantom{R\uparrow}~~\phantom{t_5} & 0.76 & [0.49,\,0.98] &  \\
 &  &  &  & B & \phantom{T\uparrow}~~\phantom{R\uparrow}~~\phantom{t_5} & 0.69 & [0.49,\,0.86] &  \\
KIC 9970396 & 1.4 & 0.7 & -3.4 & A & \phantom{T\uparrow}~~\phantom{R\uparrow}~~\phantom{t_5} & 1.06 & [0.89,\,1.24] &  \\
 &  &  &  & B & \phantom{T\uparrow}~~\phantom{R\uparrow}~~\phantom{t_5} & 1.16 & [0.92,\,1.44] &  \\
ASAS J085002+1752.5 & 1.4 & 1.1 & -3.8 & A & \phantom{T\uparrow}~~\phantom{R\uparrow}~~\phantom{t_5} & 0.99 & [0.77,\,1.24] &  \\
 &  &  &  & B & \phantom{T\uparrow}~~\phantom{R\uparrow}~~\phantom{t_5} & \nodata & \nodata &  \\
KIC 4054905 & 1.4 & 0.8 & -3.6 & A & \phantom{T\uparrow}~~\phantom{R\uparrow}~~\phantom{t_5} & 1.02 & [0.88,\,1.16] &  \\
 &  &  &  & B & \phantom{T\uparrow}~~\phantom{R\uparrow}~~\phantom{t_5} & \nodata & \nodata &  \\
KIC 5113053 & 1.3 & 0.5 & -3.3 & A & \phantom{T\uparrow}~~\phantom{R\uparrow}~~\phantom{t_5} & 0.91 & [0.56,\,1.43] &  \\
 &  &  &  & B & \phantom{T\uparrow}~~\phantom{R\uparrow}~~\phantom{t_5} & 0.93 & [0.66,\,1.32] &  \\
EPIC 203929178 & 1.3 & 0.1 & -3.0 & A & \phantom{T\uparrow}~~\phantom{R\uparrow}~~\phantom{t_5} & \nodata & \nodata &  \\
 &  &  &  & B & \phantom{T\uparrow}~~\phantom{R\uparrow}~~\phantom{t_5} & 1.12 & [0.86,\,1.36] &  \\
AD Boo & 0.8 & 0.4 & -3.8 & A & \phantom{T\uparrow}~~\phantom{R\uparrow}~~\phantom{t_5} & \nodata & \nodata &  \\
 &  &  &  & B & \phantom{T\uparrow}~~\phantom{R\uparrow}~~\phantom{t_5} & 0.98 & [0.79,\,1.20] &  \\
BW Aqr & 0.5 & 0.9 & -4.5 & A & \phantom{T\uparrow}~~\phantom{R\uparrow}~~\phantom{t_5} & 0.94 & [0.58,\,1.37] &  \\
 &  &  &  & B & \phantom{T\uparrow}~~\phantom{R\uparrow}~~\phantom{t_5} & \nodata & \nodata &  \\
RW Lac & 0.5 & 0.1 & -3.8 & A & \phantom{T\uparrow}~~\phantom{R\uparrow}~~\phantom{t_5} & 0.96 & [0.55,\,1.43] &  \\
 &  &  &  & B & \phantom{T\uparrow}~~\phantom{R\uparrow}~~\phantom{t_5} & 0.95 & [0.65,\,1.30] &  \\
Kepler-34 & 0.5 & 0.2 & -3.9 & A & \phantom{T\uparrow}~~\phantom{R\uparrow}~~\phantom{t_5} & 1.05 & [0.69,\,1.45] &  \\
 &  &  &  & B & \phantom{T\uparrow}~~\phantom{R\uparrow}~~\phantom{t_5} & 1.03 & [0.70,\,1.37] &  \\
UX Men & 0.5 & 0.2 & -3.9 & A & \phantom{T\uparrow}~~\phantom{R\uparrow}~~\phantom{t_5} & 1.02 & [0.67,\,1.46] &  \\
 &  &  &  & B & \phantom{T\uparrow}~~\phantom{R\uparrow}~~\phantom{t_5} & 1.00 & [0.66,\,1.35] &  \\
CQ Ind & 0.4 & 0.2 & -4.0 & A & \phantom{T\uparrow}~~\phantom{R\uparrow}~~\phantom{t_5} & \nodata & \nodata &  \\
 &  &  &  & B & \phantom{T\uparrow}~~\phantom{R\uparrow}~~\phantom{t_5} & 1.04 & [0.84,\,1.25] &  \\
LV Her & 0.3 & 0.2 & -4.1 & A & \phantom{T\uparrow}~~\phantom{R\uparrow}~~\phantom{t_5} & \nodata & \nodata &  \\
 &  &  &  & B & \phantom{T\uparrow}~~\phantom{R\uparrow}~~\phantom{t_5} & 1.04 & [0.64,\,1.46] &  \\
CN Lyn & 0.2 & 0.2 & -4.2 & A & \phantom{T\uparrow}~~\phantom{R\uparrow}~~\phantom{t_5} & 1.05 & [0.66,\,1.47] &  \\
 &  &  &  & B & \phantom{T\uparrow}~~\phantom{R\uparrow}~~\phantom{t_5} & 0.95 & [0.54,\,1.37] &  \\
\hline
Procyon & 16.4 & 0.3 & 14.7 & A & \phantom{T\uparrow}~~\phantom{R\uparrow}~~\phantom{t_5} & > 1.34 &  &  \\
$\mu$ Cas & 3.6 & 0.0 & 2.2 & A & \phantom{T\uparrow}~~\phantom{R\uparrow}~~\phantom{t_5} & 0.79 & [0.65,\,0.92] &  \\
\enddata
\tablecomments{Columns 2 and 3 list the $\chi^2$ of a system's isochrone fits using either standard $\solaramlt$ models or models that allow $\amlt$ to vary.  $\Delta\mathrm{BIC} = \mathrm{BIC}_{\solaramlt} - \mathrm{BIC}_{\amlt}$ offers a comparison that accounts for the differing degrees of freedom between the two fits.  The failure modes $\tilde T \uparrow(\downarrow)$ and $\tilde R \uparrow(\downarrow)$ indicate the best fitting $\solaramlt$ model is too hot/large (cool/small).  $t_5$ indicates that the best fitting $\solaramlt$ model is unphysically old, likely due to radius inflation.  Columns 7 and 8 gives best estimates of $\amlt$ inferred from the variable $\amlt$ fits; either the median and 95\% smallest credible interval, a bound, or no estimate if the $\amlt$ posterior was too unconstrained.  The last column marks cases where the 95th percentile of $|\amltA - \amltB| > 0$, indicating that $\amltA \neq \amltB$.
\newline$(*)$ using masses and radii from \citet{Kervella_2016, Kervella_2017}
\newline$(\dagger)$ using masses and radii from \citet{Akeson_2021}
\newline$(I)$ classified as magnetically inflated (see Section \ref{subsection:fast_rotators})}
\end{deluxetable*}

\subsection{Fast Rotators}
\label{subsection:fast_rotators}
It is well known that many cool stars in eclipsing binaries display radius inflation \citep{Torres_2002, Lopez-Morales_2005, Torres_2006, Lopez-Morales_2007, Cruz_2018}.  Stellar rotational period correlates with magnetic activity in convective stars \citep{Skumanich_1972, Noyes_1984}, and so the common wisdom is that the rapid rotation of close binaries drives a dynamo and strong magnetic fields that inhibit the motion of convective eddies in the outer envelope, leading to an inflated radius and lower temperature.  The mixing length parameter controls the convective efficiency and so lowering it is an ad hoc way of capturing the effect of strong magnetic fields \citep{Torres_2006, Chabrier_2007, Macdonald_2010, Vos_2012}.  

The correlation between rotation and magnetic activity is often recast using the Rossby number $\Ro = P_\mathrm{rot}/\tau_\mathrm{conv}$, a dimensionless number encoding the relative strength of convective motions against Coriolis forces.  Here we use the ``global" convective turnover time $\tau_\mathrm{conv}$ provided in the \misttwo\ tables, calculated as $\tau_\mathrm{conv} = \int_{R_{cz}}^{R_*} \frac{dr}{v_\mathrm{conv}}$ integrated over the convective envelope.  There is an observed threshold of $\Ro \lesssim 0.13$ below which stars tend to be strongly magnetic \citep{Reiners_2022}.  Most of our \debs\ have orbital periods short enough that they should be tidally synchronized \citep{Torres_2010} and so we can approximate the Rossby number as $\Ro \approx P_\mathrm{orb} / \tau_\mathrm{conv}$.  

\begin{figure}[t!]
\epsscale{1.0}
\plotone{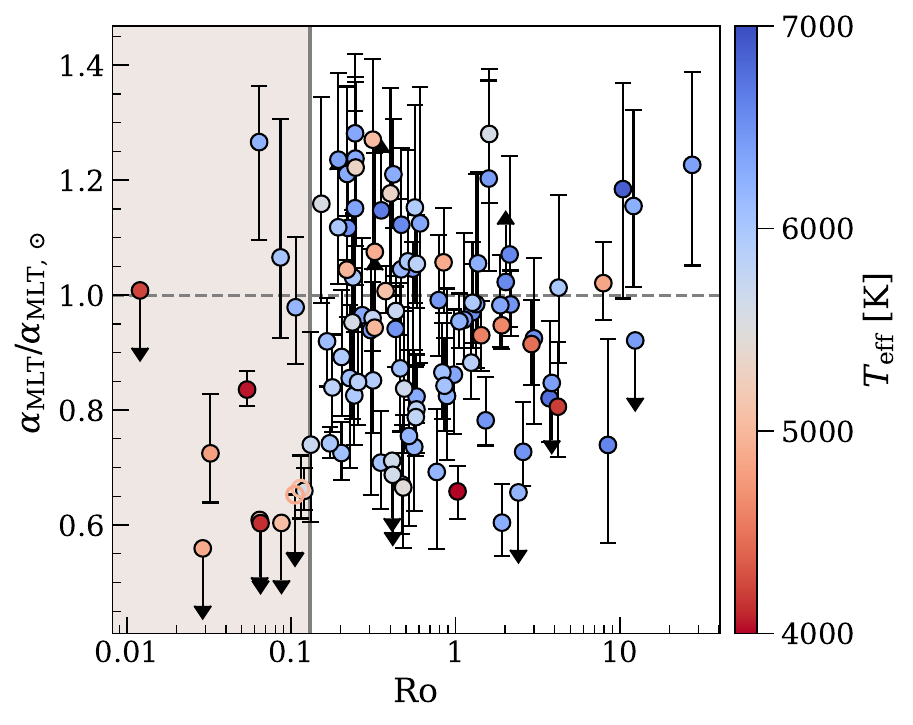}
\caption{Rossby number $\Ro$ vs. inferred $\amlt$ of the \debs.  Stars with wide, uninformative $\amlt$ posteriors are not plotted; stars with only upper/lower bounds in Table \ref{table:debs_results} are plotted with arrows.  An age prior of $t~ < 13.8$ Gyr was applied to TIC 172900988 and so it's components are plotted as open circles.  Cool stars ($\Teff \lesssim 5400$ K) that are rapidly rotating ($\Ro < 0.13$; shaded region) are generically found to have low $\amlt$.  This is the familiar rotation + magnetically induced radius inflation seen in K dwarfs \citep{Lopez-Morales_2007}.\label{figure:debs_amlt_vs_Ro}}
\end{figure}

Figure \ref{figure:debs_amlt_vs_Ro} plots $\Ro$ versus the inferred $\amlt$ of the \debs\ and we can clearly see that stars who are both cool ($\Teff < 5400$ K) and rapidly rotating ($\Ro < 0.13$) generically have low mixing lengths.  All but one of these stars (RU Cnc B) are flagged with some combination of the $\tilde T \uparrow$, $\tilde R \downarrow$, or $t_5$ flags in Table \ref{table:debs_results}, consistent with radius inflation.  We mark all nine of these stars as ``inflated" in Table \ref{table:debs_results} and do not consider them in any of the subsequent analyses or figures.

\subsection{$\amlt$ versus Stellar Parameters}
\label{subsection:}
\begin{figure}[h!]
\epsscale{1.0}
\plotone{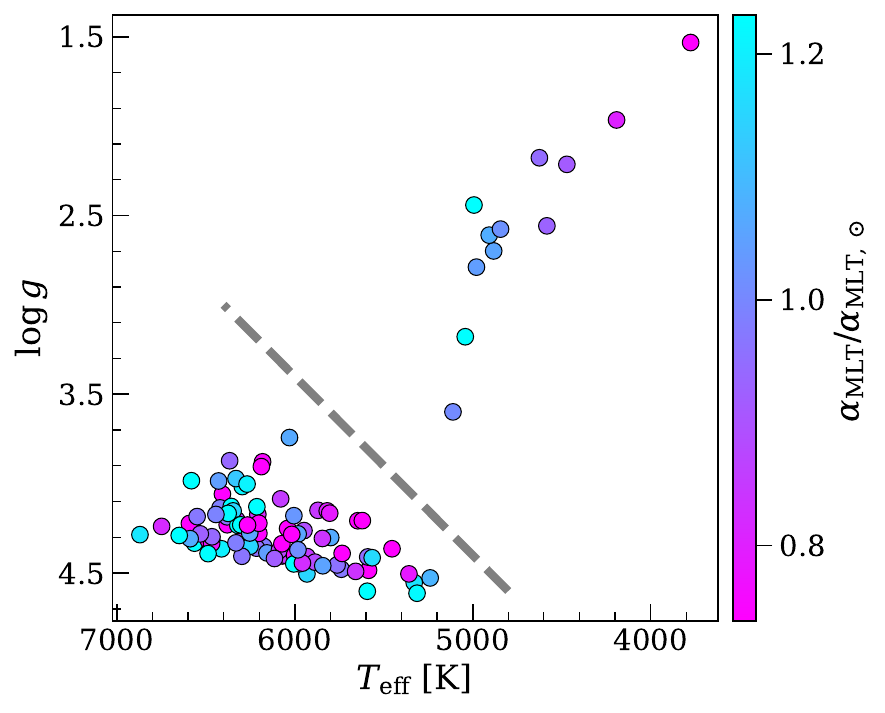}
\caption{DEBs in the Kiel diagram colored by inferred $\amlt/\solaramlt$.  Again, rotationally inflated stars and stars with uninformative $\amlt/\solaramlt$ posteriors are not plotted.  We can separate the sample into two groups of `evolved' and `main sequence' stars.  The evolved stars show a strong correlation between $\amlt/\solaramlt$ and Kiel position, with mixing length decreasing up the giant branch.\label{figure:debs_kiel_amlt}}
\end{figure}

Simulations and other empirical studies have both previously found correlations between mixing length and stellar parameters, suggesting the possibility for a simple calibration of $\amlt$ in stars beyond the Sun.  Here we investigate whether these correlations are borne out in our own \debs\ sample.  Figure \ref{figure:debs_kiel_amlt} plots the \debs\ in Kiel space colored by their inferred $\amlt/\solaramlt$.  The sample separates cleanly into two groups, evolved stars on the giant branch and less evolved, mostly main-sequence stars.  The main-sequence stars show large scatter in their mixing lengths, but there appears to be a trend where $\amlt/\solaramlt$ steadily decreases with decreasing $\logg$ on the giant branch.  

\begin{figure*}[]
\epsscale{0.8}
\plotone{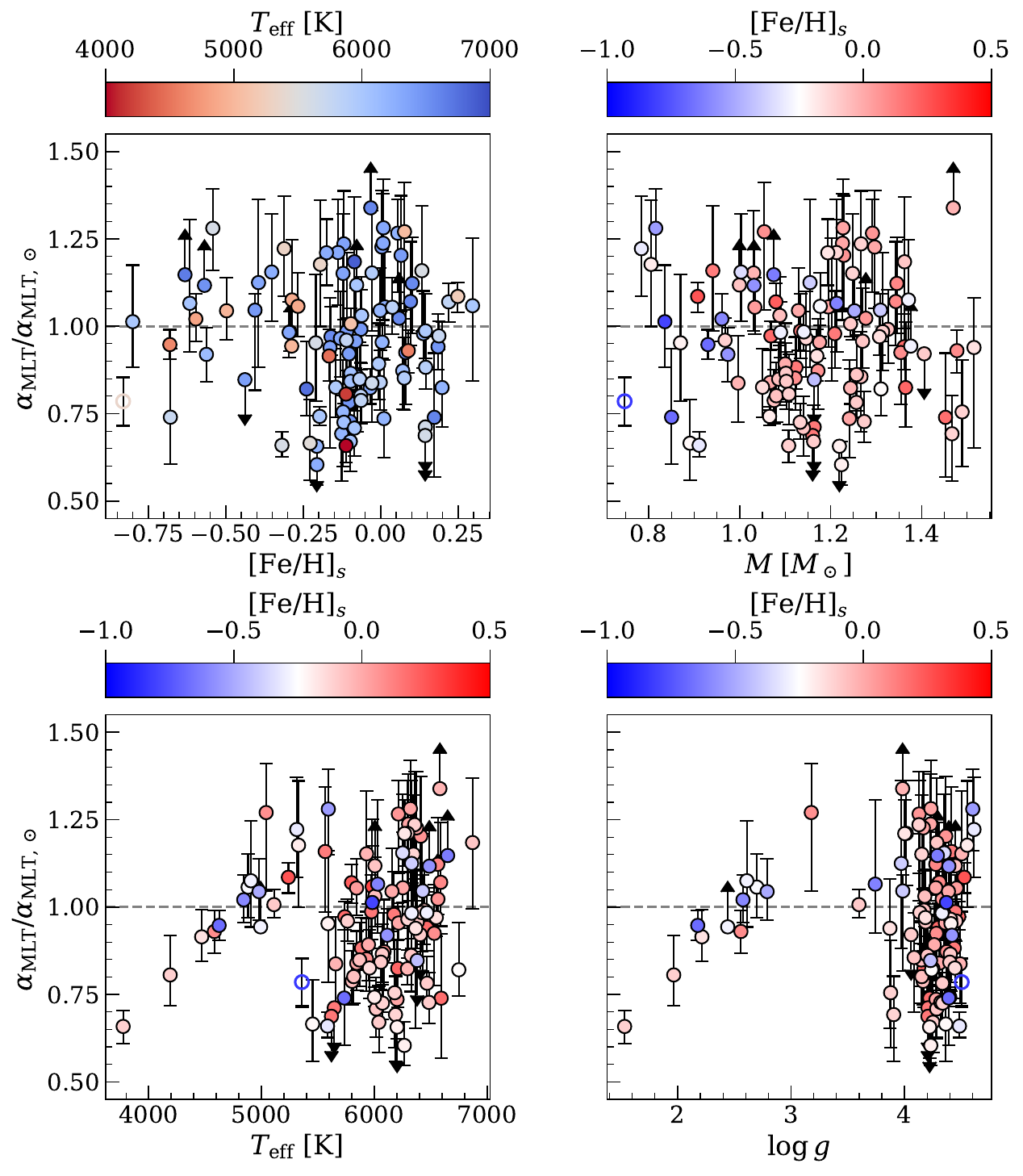}
\caption{Inferred $\amlt/\solaramlt$ vs stellar parameters for the DEBs.  There appears to be a correlation between $\amlt/\solaramlt$ and $\logg$ or $\Teff$ on the giant branch (see text) and we do not find the mixing lengths of main-sequence DEBs to systematically vary with stellar parameters, contrary to previous empirical correlations observed in asteroseismic samples \citep{Bonaca_2012, Metcalfe_2014, Creevey_2017, Viani_2018, Li_2024}.\label{figure:debs_amlt_vs_stellar_parameters}}
\end{figure*}

Figure \ref{figure:debs_amlt_vs_stellar_parameters} plots $\amlt/\solaramlt$ versus the stellar parameters $\FeHs$, mass, $\Teff$, and $\logg$.  Here we see more clearly the correlation between mixing length and position along the giant branch.  There's a seemingly tight correlation between $\amlt/\solaramlt$ and both $\Teff$ and $\logg$ for stars with $\logg \lesssim 3.5$, where $\amlt/\solaramlt$ decreases at lower temperatures and surface gravities, i.e., higher up on the branch.  Such a correlation implies a systematic offset in the temperature scale of $\solaramlt$ models on the giant branch compared to the observed temperatures.  However, most of the $\chi^2_\odot$ values on the giant branch in Figure \ref{figure:debs_solaramlt_kiel} are low and do not suggest that $\solaramlt$ models are performing poorly in this regime.  The actual temperature residuals of the $\solaramlt$ RGB models are noisy, only slightly correlated with $\logg$, and importantly all but one star have residuals $|\Teff - \tilde T_\mathrm{eff} (\hat{\boldsymbol\theta})| < 1\sigma_{\Teff}$.  Thus we conclude that the apparently strong $\amlt$ correlation on the giant branch is not actually significant and that the strong sensitivity $\partial\Teff / \partial\amlt$ on the branch is causing $\sim1\sigma$ temperature residuals to appear like an overly dramatic variation in $\amlt$.

The lack of other correlations with stellar parameters on the giant branch, particularly metallicity, is interesting.  \citet{Li_2024} studied a sample of oscillating Kepler giants, a handful of which were giant-giant binaries displaying both oscillations and eclipses.  Modeling the oscillation frequencies together with the dynamical constraints from eclipses they found a relatively steep, positive correlation between mixing length and metallicity, with $\amlt/\solaramlt \propto 0.3 \times \mathrm{[M/H]}$.  
We do not observe such an obvious correlation with metallicity on the giant branch and instead observe $\amlt/\solaramlt$ to only depend on $\Teff$ or $\logg$.  \citet{Li_2024} do however highlight that the choice of terms included in the likelihood is important, with different combinations of seismic and dynamical constraints altering the correlations they observe between $\amlt/\solaramlt$ and stellar parameters.

The aforementioned literature correlations between $\amlt$ and stellar parameters have often been cast in a simple trilinear relationship of the form  $\amlt/\solaramlt = a + b\times\logg + c\times\logTeff + d\times\FeH$ \citep[e.g.,][]{Bonaca_2012}.  We did not find by eye any obvious trends between the $\amlt$ and stellar parameters among the main-sequence \debs, but here we test against particular trilinear relations reported in the literature.  These sorts of calibrations are likely to find use in future stellar analysis pipelines and so it is prudent to test if they are consistent with the constraints offered by eclipsing binaries.  We specifically test the empirical $\amlt$ relation \citet{Viani_2018} observed in solar-like oscillators and the $\amlt$ calibration of \citet{Magic_2015}, informed by the STAGGER grid of 3D RHD simulations of stellar surface convection \citep{Magic_2013}.  

\begin{figure*}[t!]
\epsscale{1.0}
\plotone{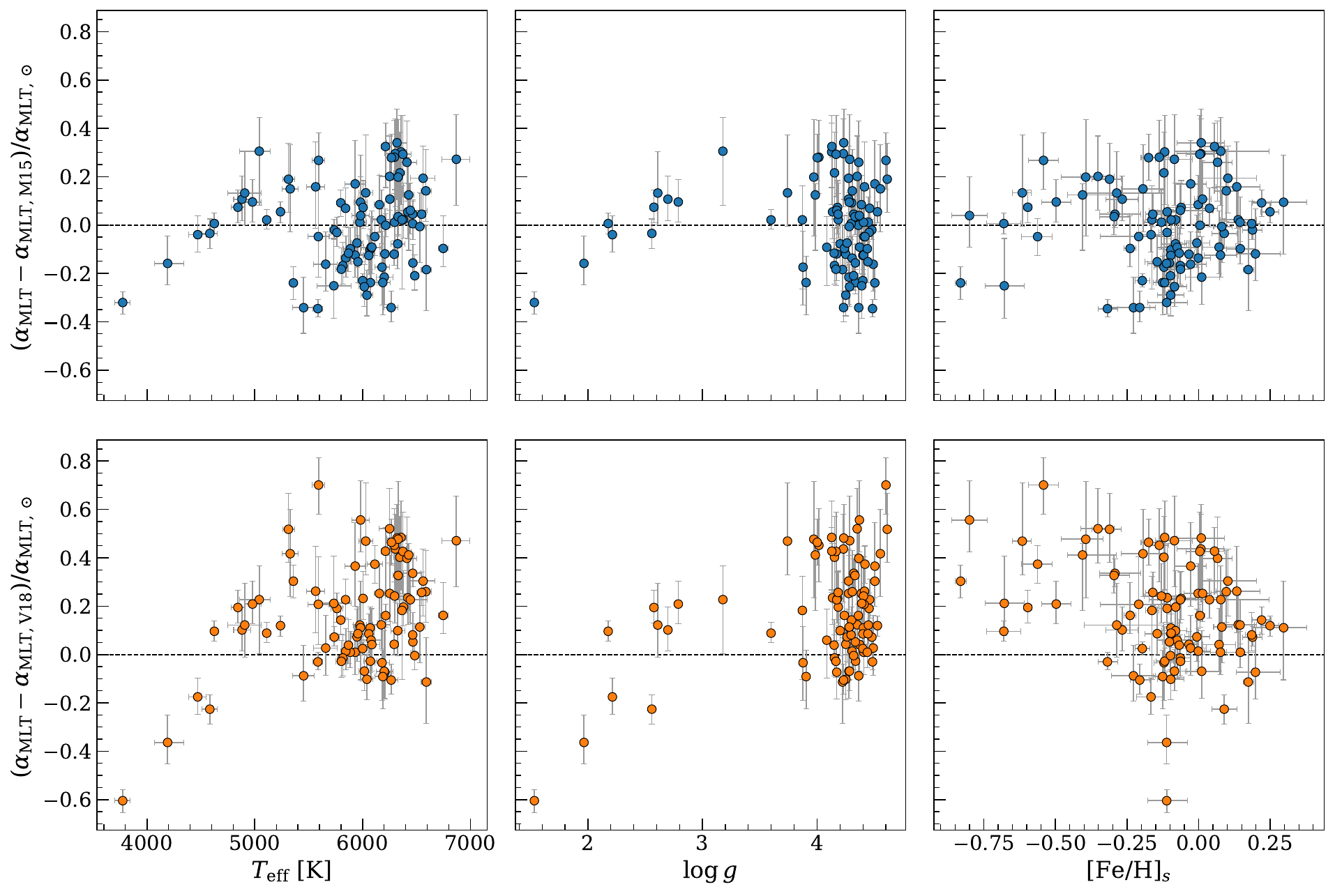}
\caption{Residuals between our inferred mixing lengths and those predicted by the trilinear calibrations of \citet{Magic_2015} (M15; top) and \citet{Viani_2018} (V18; bottom).  These $\amlt$ calibrations from the literature are wholly inconsistent with the mixing lengths we infer in the \debs\ and so they do nothing to decrease the scatter.
\label{figure:debs_literature_trilinear_residuals}}
\end{figure*}

In Figure \ref{figure:debs_literature_trilinear_residuals} we plot the residuals of our inferred $\amlt$ versus that predicted by these two trilinear prescriptions.  \citet{Magic_2015} predicts only mild variation in $\amlt$ away from the solar-calibrated value so the residuals look similar to the absolute $\amlt/\solaramlt$ scatter in Figure \ref{figure:debs_amlt_vs_stellar_parameters}.  The \citet{Viani_2018} trilinear relation predicts much stronger variation in $\amlt$, particularly with metallicity, which we can see imparted into the residuals when we subtract its predictions from our own mixing lengths.  The standard deviation of the residuals $\sigma(\Delta)$ with respect to the \citet[][M15]{Magic_2015} prediction is $\sigma(\Delta) = 0.177$, and $\sigma(\Delta) = 0.194$ for \citet[][V18]{Viani_2018}.  These can be compared to the null standard deviation of the \debs' $\amlt/\solaramlt$ distribution, $\sigma(\amlt/\solaramlt) = 0.173$.  Subtracting off these literature predictions does nothing to reduce the scatter in $\amlt/\solaramlt$ and thus they are inconsistent with the constraints provided by the \debs.  We also attempted to fit our own trilinear relationship to our inferred mixing lengths, taking into account the correlations between binary components, but we found that removing the resulting fitted trend also did not reduce the scatter in any meaningful way.  We conclude that there is no statistically robust correlation between $\amlt$ and $\Teff$, $\logg$, or $\FeH$ within the \debs\ sample.

\begin{figure}[t!]
\epsscale{1.0}
\plotone{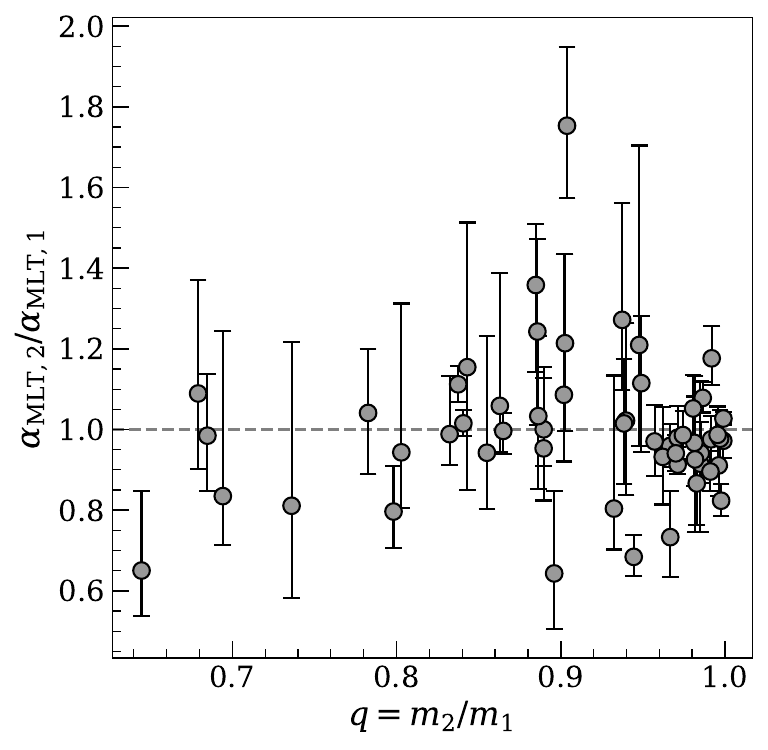}
\caption{Ratio between component mixing lengths ${\amlt}_2/{\amlt}_1$ and the mass ratio $q$ in \debs.  The systems scatter evenly around ${\amlt}_2/{\amlt}_1 = 1$ down to the lowest mass ratios, indicating that there is no global mass ordering observed between the component mixing lengths in our binaries.\label{figure:debs_amlt_ratio_vs_mass_ratio}}
\end{figure}

It has also been suggested that the mixing length in main-sequence stars might depend on stellar mass, though there are claims of both a positive \citep{Yildiz_2006, Kervella_2008} and negative \citep{Joyce_2018a, Joyce_2023a} sign to the dependence.  Our \debs\ offer a powerful test to any hypothetical correlation with mass because we jointly fit both components in a binary simultaneously, each with their own mixing length.  If $\amlt$ does depend on mass then we would expect the joint $(\amltA, ~\amltB)$ posterior to indicate some preferred mass ordering.  We plot the ratio of component mixing lengths ${\amlt}_2/{\amlt}_1$ versus the binary's mass ratio $q = m_2/m_1$ of the main-sequence binaries in Figure \ref{figure:debs_amlt_ratio_vs_mass_ratio}.  Here the more massive star in each binary is labeled as ``1" so $m_1 \geq m_2$ and $q \leq 1$.  We find that the mixing length ratios scatter evenly above and below ${\amlt}_2/{\amlt}_1 = 1$, indicating that there is no systematic preference for the mixing lengths to be mass ordered within a binary pair.  

\subsection{Individual Cases of Non-solar Mixing Length}
\label{subsection:case_studies}
Though we do not find any robust variation in mixing length as a function of stellar parameters we can still identify several particular binary systems that are both poorly fit by $\solaramlt$ models and better fit by allowing for a non-solar mixing length.  In this section we highlight the most interesting cases that are good candidates for more detailed case studies in the future.  This discussion is not strictly exhaustive of all of the interesting cases so we again refer the reader to Table \ref{table:debs_results} for a complete census of our isochrone fits to the \debs.

\subsubsection{$\alpha$ Centauri}
\label{subsection:discuss_alphaCen}
$\alpha$ Centauri (HIP 71683 \& HIP 71681) is our nearest neighboring star system and has long been one of the most crucial testbeds for stellar models outside of our own Sun \citep{Lattanzio_1984, Demarque_1986, Brown_1994, Fernandes_1995, Miglio_2005, Yildiz_2007, Joyce_2018a, Nsamba_2018} and so we offer it special attention here.  Recently, \citet{Joyce_2018a} leveraged the interferometric radii of \citet{Kervella_2017}, $p-$mode oscillations from \citet{Kjeldsen_2005} and \citet{deMeulenaer_2010}, and the dynamical masses of \citet{Kervella_2016} to infer the mixing length in both components and found that these observations are inconsistent with a solar mixing length.  They instead found consistency only when $\amltA < \alpha_\odot$ and $\amltB > \alpha_\odot$, though the precise values of $\amltA$ and $\amltB$ were strongly degenerate with the system age and this degeneracy was only broken by the inclusion of the asteroseismic measurements.  The relative ordering of $\amltA < \amltB$ was robust however, being present at every age even without the inclusion of the seismic data and persisting across different input physics assumptions (e.g., atomic diffusion, atmospheric boundary condition, core overshooting).

\begin{figure}[t!]
\epsscale{1.0}
\plotone{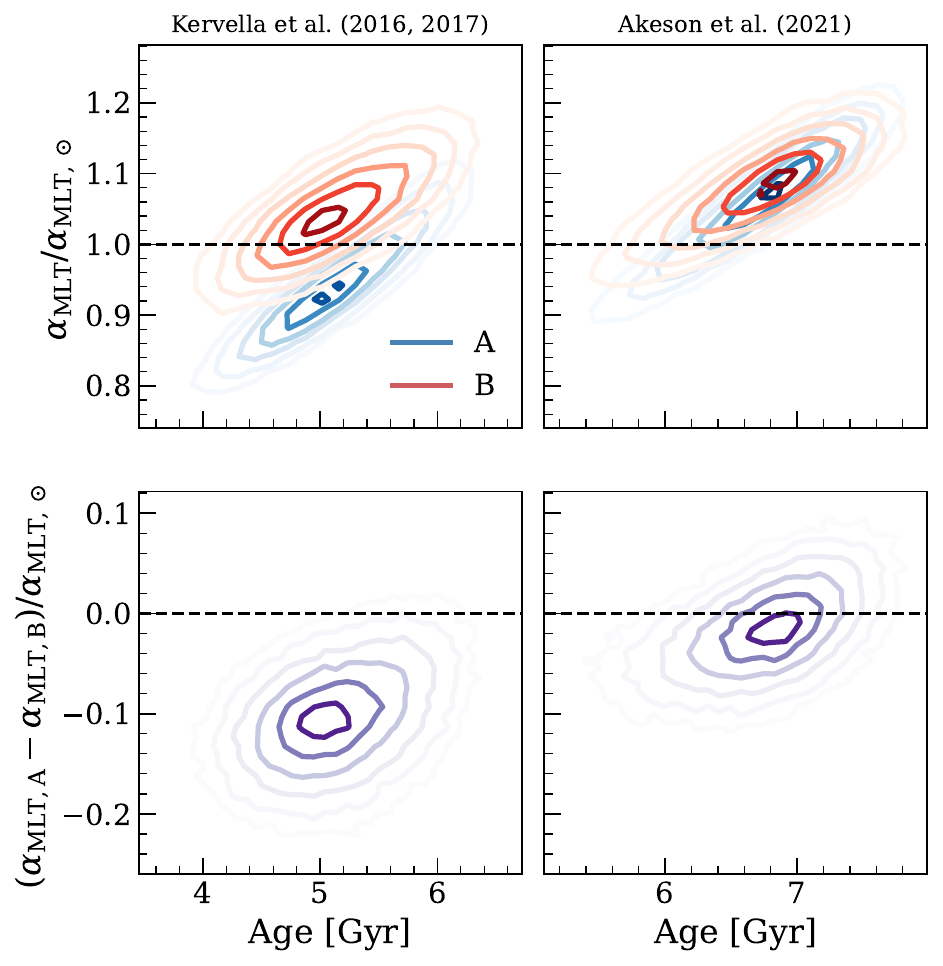}
\caption{Age$-\amlt$ posteriors for $\alpha$ Centauri A \& B, using either the masses and radii of \citet{Kervella_2016, Kervella_2017} (left) or \citet{Akeson_2021} (right).  Contour lines are drawn at every $0.5\sigma$ confidence interval up to $3\sigma$.  Using the \citet{Kervella_2016, Kervella_2017} measurements it is clear that requiring a common age for both binary components implies $\amltA \neq \amltB$ and therefore one or both components' mixing lengths must differ from $\solaramlt$.  The two components' mixing lengths are similar to each other using the \citet{Akeson_2021} measurements but still prefer $\amlt/\solaramlt\approx 1.1$.\label{figure:discuss_alphaCen}}
\end{figure}

When using the \citet{Kervella_2016, Kervella_2017} masses and radii our variable $\amlt$ run reaches an identical conclusion to \citet{Joyce_2018a}, with $\amltA < \amltB$ across all ages at high significance (left of Figure \ref{figure:discuss_alphaCen}).  Given that our $\amlt = \solaramlt$ run was flagged as having a high $\chi^2$ and poorly matching $R_B$ it would appear that a non-solar mixing length is necessary to properly model $\alpha$ Centauri.  

However, \citet{Akeson_2021} used new absolute astrometry in the millimeter to re-solve the system's orbital scale, making a slight revision to the radii and measuring updated masses that are $\sim5\sigma$ discrepant with the earlier \citet{Kervella_2016} measurements.  When using these newer \citet{Akeson_2021} measurements our isochrone fits produce qualitatively different results than the runs that used the \citet{Kervella_2016, Kervella_2017} measurements; the $\solaramlt$ models now provide a reasonable fit with a $\chi^2 = 4.6$, and the variable $\amlt$ run infers $\amltA \approx \amltB$ at most ages with a moderate preference for $\amlt/\solaramlt \approx 1.1$ (right of Figure \ref{figure:discuss_alphaCen}).  The previously strong signal of a non-solar mixing length in both $\alpha$ Cen A \& B is made more tenuous when the updated \citet{Akeson_2021} measurements are adopted, though a non-solar mixing length is still supported by the data.  A re-analysis of $\alpha$ Cen like that of \citet{Joyce_2018a} but instead using the \citet{Akeson_2021} masses would be very revealing.

\subsubsection{Procyon}
\label{subsection:discuss_Procyon}
\begin{figure}[h]
\epsscale{1.0}
\plotone{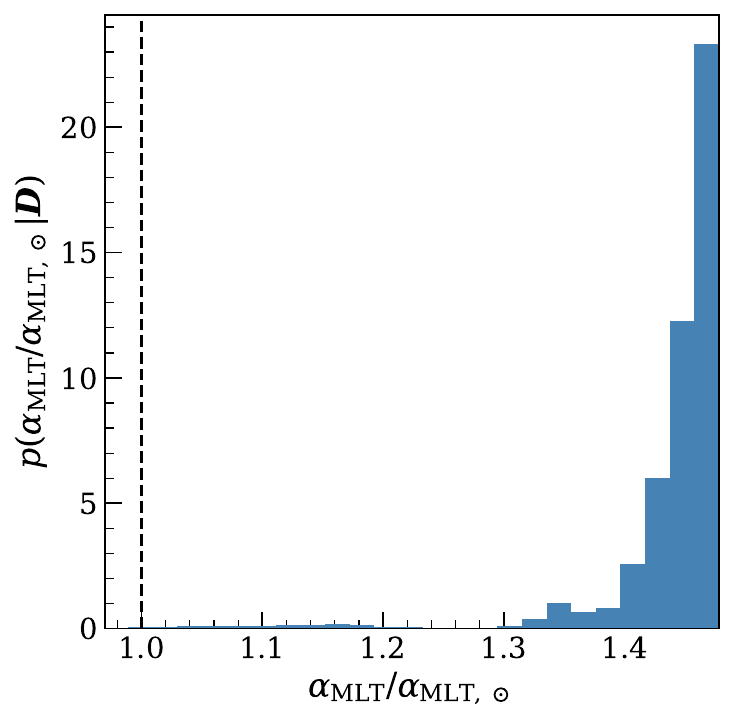}
\caption{Marginal posterior of absolute mixing length $\amlt$ in Procyon A.  The dynamical mass and angular radius measurement come from \citet{Bond_2015} and \citet{Baines_2021} respectively, while the spectroscopic parameters come from \citet{Soubiran_2024, Casamiquela_2026}.  The inferred $\amlt$ is slammed up against the upper prior edge, giving only a lower limit.  Procyon A's measurements do require non-standard stellar modeling, but $\amlt$ is poorly suited as the free parameter to tune in this particular case (see text).\label{figure:Procyon_amlt_hist}}
\end{figure}

Procyon ($\alpha$ CMi, HIP 37279) is another nearby visual binary with a long history as a non-solar calibrator \citep{Brown_1991, Irwin_1992, Guenther_1993, Chaboyer_1999, Kervella_2004, Robinson_2005, Guenther_2014, Bond_2015, Moedas_2026}.  Our variable $\amlt$ fit to Procyon A effectively only infers a lower limit as the $\amlt$ posterior is slammed up against the upper prior edge (Figure \ref{figure:Procyon_amlt_hist}).  The $\solaramlt$ fit to Procyon actually reproduces the very precise observed radius and temperature but at the cost of severely overestimating the mass $\tilde M (\hat{\boldsymbol\theta}) - M \approx 4\sigma_{M}$.  In other words the $M-L$ relation predicted by standard models is inconsistent with the observations of Procyon, with it being seemingly over-luminous given its mass.  Mixing length primarily controls the main sequence $M-\Teff$ and $M-R$ scales in a way that preserves luminosity (see Figure \ref{figure:amlt_effect}) and so it is simply not an effective model parameter to tune in order to fix this large discrepancy.  The inferred mixing length must therefore slam up against the prior edge in order to make any headway in mitigating the mass residual.

Bespoke modeling of Procyon instead typically relies on tuning the convective core overshooting.  Models of Procyon have consistently required large amounts of overshooting in order to match the classical and seismic observations \citep{Guenther_2014, Bond_2015}.  The seismic frequencies in particular are discerning because they better probe the size of the convective core itself, so the high overshooting is more than just a statement about total stellar radius.  Our models have only a fixed core overshooting and so we interpret the high $\amlt$ in this case as trying to make up for the insufficient level of overshooting, implying that the $\amlt$ signal here is not genuine.

\subsubsection{$\mu$ Cassiopeia}
\label{subsection:discuss_muCas}
$\mu$ Cassiopeia (HIP 5336) is a metal-poor \citep[$\FeHs = -0.8$;][]{Jofre_2015, Casamiquela_2026} visual binary comprised of a low mass dwarf ($M = 0.74M_\odot$) and a faint dM dwarf companion.  \citet{Bond_2020} used ground based observations and HST astrometry to solve the system's orbit and component masses, and \citet{Boyajian_2008} resolved the angular diameter of the primary star with the CHARA array.  $\mu$ Cas A is also a Gaia FGK benchmark star \citep{Soubiran_2024} and thus very well characterized spectroscopically. 

\begin{figure}[h]
\epsscale{1.0}
\plotone{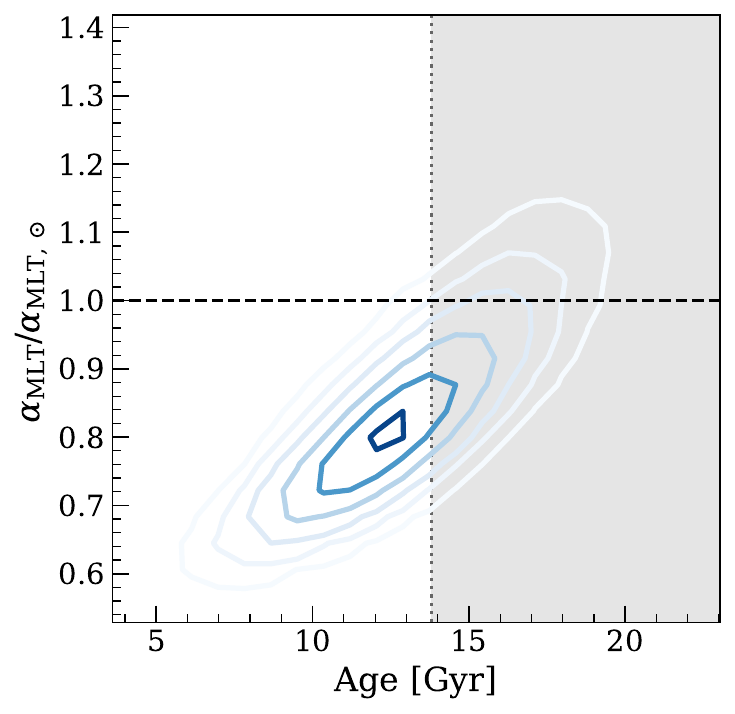}
\caption{Age$-\amlt/\solaramlt$ posterior for the metal-poor dwarf $\mu$ Cassiopeia A.  The dynamical mass and angular radius measurement come from \citet{Bond_2020} and \citet{Boyajian_2008}, while the spectroscopic parameters come from \citet{Soubiran_2024}.  Only models with $\amlt < \solaramlt$ are capable of reproducing the observations, particularly the radius and temperature, at physically realistic ages.\label{figure:muCas_amlt_age}}
\end{figure}

As mentioned in the discussion of Figure \ref{figure:debs_solaramlt_kiel} our $\solaramlt$ fit to $\mu$ Cas A returned an age that is $\approx 1.5\sigma$ greater than the accepted age of the universe, suggesting some mild issue with our standard $\solaramlt$ models.  The variable $\amlt$ fit is very revealing; the covariance between $\amlt$ and age in the inference implies that solutions with physically realistic ages require $\amlt < \solaramlt$.  This can clearly be seen in Figure \ref{figure:muCas_amlt_age} where we plot the age$-\amlt$ posterior for $\mu$ Cas A.  

The rotation period of $\mu$ Cas A is not known but it is clear that the star is very old, and being in a well separated binary it is free of tidal forces, so there is no expectation of rapid rotation that could induce magnetic inflation.  \citet{Bond_2020} note that the angular diameter of $\mu$ Cas A is close to the resolution limit of $K$ band interferometers which could be driving a systematic bias in the angular radius.  A downward revision of the angular radius relative to \citet{Boyajian_2008} could neatly bring $\solaramlt$ models back into alignment at realistic ages.  

If instead the current measurements are truly sound then we must conclude that standard stellar models fail to properly predict the radius of this metal-poor dwarf.  A sub-solar mixing length is a very plausible way to explain this discrepant radius and we note that a sub-solar mixing length at low metallicity is in qualitative agreement with the metallicity dependence reported by other empirical mixing length calibrations \citep{Bonaca_2012, Metcalfe_2014, Viani_2018, Joyce_2018, Li_2024} as well as in agreement with our findings in the \interferometric\ sample (Section \ref{section:irad_results}).   Given the diagnostic potential of $\mu$ Cas A for low metallicity stellar models we agree with \citet{Bond_2020} that the radius of $\mu$ Cas A should be resolved by instruments at shorter wavelengths and higher spatial resolutions in order to clarify the question of its discrepant radius.

\subsubsection{Cases where $\amltA \neq \amltB$}
\label{subsection:debs_results_relative_amlt}

\begin{figure*}
\epsscale{1.0}
\plotone{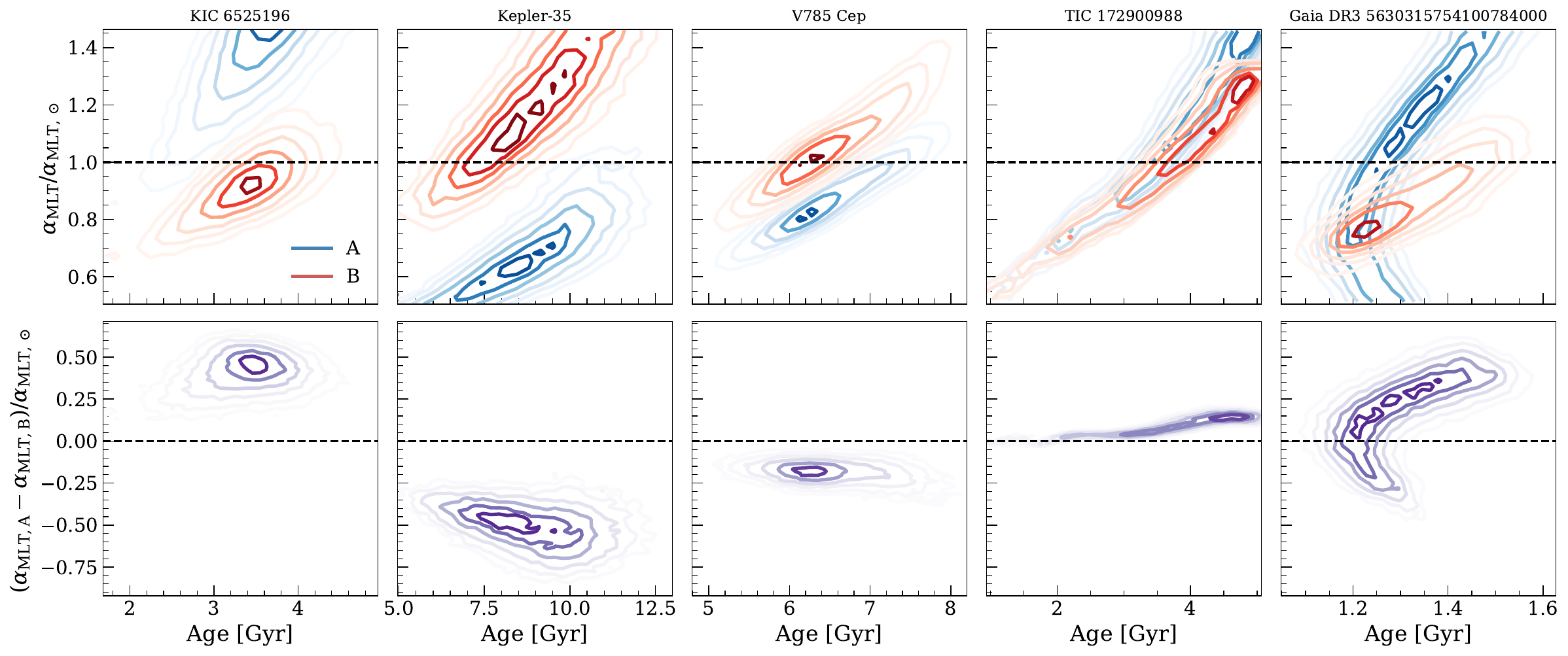}
\caption{Individual $\amlt$ and $\amltA - \amltB$ versus age posteriors for systems poorly fit by $\solaramlt$ models and flagged with $\amltA \neq \amltB$ in Table \ref{table:debs_results}.  We fit both components of each binary simultaneously with independent mixing lengths and so the covariance between $\amltA$ and $\amltB$ can ensure that $\amltA \neq \amltB$ at every age, indicating that at least one component must have a non-solar mixing length.\label{figure:debs_unequal_amlt}}
\end{figure*}

There are fives cases in Table \ref{table:debs_results} that have $\chi^2_\odot > 9.49$ ($\sim 2\sigma$ discrepant), $\chi^2_\odot > \chi^2_{\amlt}$, and are flagged as $\amltA \neq \amltB$.  Figure \ref{figure:debs_unequal_amlt} plots these systems' marginal $\amlt$ and $\amltA - \amltB$ versus age posteriors in order of decreasing $\chi^2_\odot$.  These systems are similar to the \citet{Kervella_2016, Kervella_2017} case for $\alpha$ Cen (left of Figure \ref{figure:discuss_alphaCen}) where the \emph{joint} posterior between $\amltA$ and $\amltB$ reveals a non-solar mixing length signal, even if the marginal mixing length posteriors when considered individually may appear consistent with the solar-calibrated value.

The $\solaramlt$ fit to KIC 6525196's primary star severely underpredicts the mass while successfully reproducing the radius and temperature, suggesting discrepancy in the $M-L$ relation of standard models, similar to Procyon A (Section \ref{subsection:discuss_Procyon}).  The primary's $\amlt$ posterior is also pushed against the upper prior edge similar to Procyon A, though far less dramatically.  The variable $\amlt$ models do manage to achieve a reasonably good fit to the system overall but there is still some room for improvement, and given the primary's posterior hitting the prior edge we feel that this system is worth further investigation.  This system is also metal-poor ($\FeH\approx -0.5$) and so may be revealing deficiencies in standard models of metal-poor main-sequence stars.

Kepler-35 and V785 Cep are particularly similar to the \citet{Kervella_2016, Kervella_2017} $\alpha$ Cen fits, with the primary star having a consistently lower mixing length than the secondary star across all ages.  Kepler-35 is a pair of old K dwarfs that show an incredibly strong mixing length signal.  The system was originally characterized by \citet{Welsh_2012} for the discovery of a circumbinary planet but the stellar components appear to be equally interesting.  We highly recommend Kepler-35 be explored in a future case study in order to understand why these otherwise ordinary appearing dwarfs are in tension with standard stellar models.

V785 Cep is a pair of slightly more massive ($M = 1.10M_\odot$ and $1.07M_\odot$) solar metallicity stars close to the end of their main-sequence lives, with the more massive primary being $\approx 150$ K cooler than the secondary.  The best fitting $\solaramlt$ models do a very poor job of matching the observed temperatures, instead preferring them to be equal.  The radii and masses are reproduced well in both the $\solaramlt$ and variable $\amlt$ fits and so the mixing length signal is coming almost entirely from the temperatures.  The system's temperature ratio is one of the more robust results of the light curve analysis and so the poor $\solaramlt$ fit should be reliable.  But spectroscopic temperatures can be difficult to measure reliably in an eclipsing SB2 system like this and so a reexamination of V785 Cep's spectroscopic solution would confirm the apparently constant offset between component mixing lengths.

The $\amlt$ posteriors for both components in TIC 172000988 are relatively poorly constrained, and the primary's is even formally classified as uninformative (Table \ref{table:debs_results}).  Despite the wide posteriors the mixing lengths inferred for TIC 172000988 tend super-solar and also follow a relative ranking of $\amltA > \amltB$ at the 95\% confidence level.  This result is admittedly very marginal and both components are right around the mass where the core transitions from being radiative to convective ($\sim 1.2M_\odot$), so the precise structure of these stars will be highly sensitive to other model physics assumptions beyond mixing length, e.g., opacities and convective core overshooting.  We therefore do not consider the $\amlt$ signal in TIC 172000988 to be particularly robust.

The last binary in this group is Gaia DR3 5630315754100784000.  Both components are intermediate mass main-sequence stars and the primary is especially warm ($\Teff \approx 7100$ K), hence the poorly constrained mixing length.  The mixing length of the secondary however is decently well constrained and tends sub-solar.  The covariance between the two mixing lengths is also interesting despite the poor constraint on the primary.  $\amltB$ is only consistent with a solar-calibrated mixing length when $\amltA$ is highly super-solar, and conversely if we insist that $\amltA = \solaramlt$ then $\amltB$ is forced to increasingly sub-solar values.  The morphology of the ($\amltA$, $\amltB$) posterior in this system is complex but it does confidently indicate some combination of non-solar mixing lengths are necessary to match the data.

\subsubsection{Other Cases of Non-solar Mixing Length}

\begin{figure*}
\epsscale{0.8}
\plotone{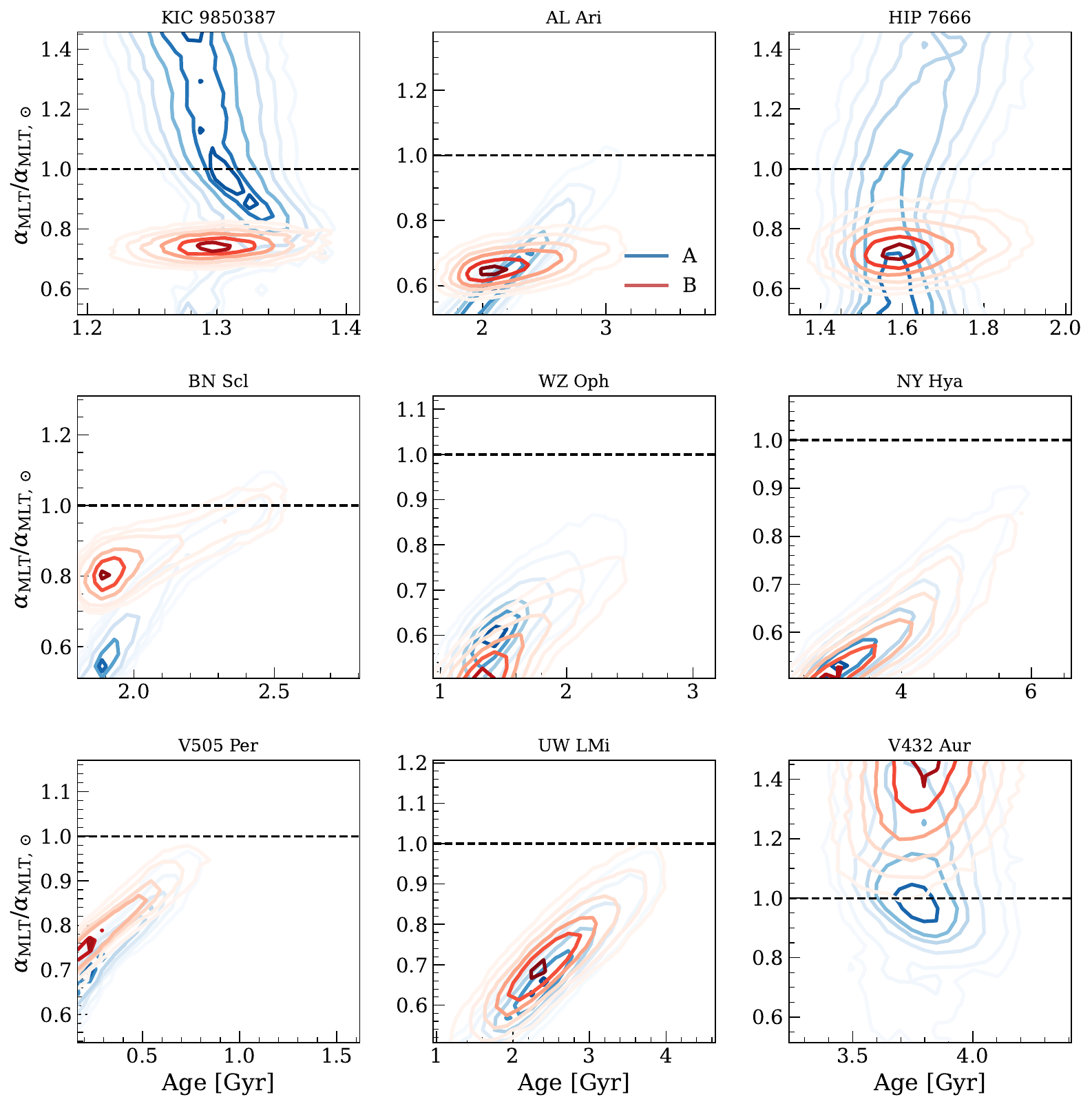}
\caption{Individual $\amlt-$age posteriors for the remaining systems that are poorly fit by $\solaramlt$ models, in order of decreasing $\solaramlt~\chi^2$.  Most of these cases (AL Ari, BN Scl, WZ Oph, NY Hya, V505 Per, UW Lmi) could be plausibly considered inflated in a fashion similar to the rapidly rotating K-type stars (Section \ref{subsection:fast_rotators}), but these system are either rotating far more slowly or are far warmer.  If magnetic inflation truly is the culprit behind their poor standard model fits then these systems offer insights into the magnetic inflation mechanism in a regime qualitatively distinct than K dwarfs.\label{figure:debs_other_nonsolar_amlt}}
\end{figure*}

Figure \ref{figure:debs_other_nonsolar_amlt} plots the $\amlt-$age posteriors for the remaining \debs\ whose variable $\amlt$ models offer significant improvement of the poor $\solaramlt$ fits.  Five of these systems, AL Ari, BN Scl, WZ Oph, NY Hya, V505 Per, and UW LMi, cleanly display sub-solar mixing lengths in both components.  This is reminiscent of the radius inflation seen in the rapidly rotating K dwarfs \ref{subsection:fast_rotators} but invoking the magnetic inflation mechanism in these cases comes with some caveats.  All stars in these five systems except for AL Ari A and NY Hya A \& B are warm ($\Teff > 6000$ K) and slowly rotating ($\Ro\sim 0.4 - 12.0$) and so the conditions of their convective envelopes are very different from those of the magnetically inflated K dwarfs.  NY Hya is known to be magnetically active \citep{Hinse_2024} and so lends support to the idea that magnetic inflation can manifest in more slowly rotating G and F-type main-sequence stars, though admittedly its components are still $>400$ K cooler than all of the other stars mentioned above and so they might not be representative.  Claiming that all of these stars are magnetically inflated also does not explain why there are other stars of similar temperatures and rotation that show no signs of magnetic inflation (Figure \ref{figure:debs_amlt_vs_Ro}).  Nevertheless, these systems are interesting cases with which to test non-standard stellar models, whether that is testing prescriptions for magnetism or exploring other non-standard model physics.

The other three systems appear to be qualitatively different.  KIC 9850387 is a binary consisting of a massive subgiant ($M = 1.66M_\odot$, $R = 2.15 R_\odot$) and a Sun-like companion ($M = 1.06M_\odot$, $R = 1.08 R_\odot$) that $\solaramlt$ models do an incredibly poor job of fitting.  The system's age is tightly constrained by the HR position of the subgiant and so we confidently infer a sub-solar mixing length in the lower mass companion.  The subgiant component displays solar-like oscillations of mixed character \citep{Sekaran_2020} and \citet{Sekaran_2021} found the dynamical and seismic solutions for this star to be in tension.  Combined with the sub-solar mixing length we infer for the secondary, KIC 9850387 proves to be a very intriguing system that can serve as a calibrator towards improved modeling of mixing within stars.  

HIP 7666 is similarly a pair of a more massive, pulsating evolved star ($M = 1.48 M_\odot$, $R = 2.04 R_\odot$) and a relatively more solar companion \citep[$M = 1.15M_\odot$, $R = 1.20 R_\odot$;][]{Escola-Sirisi_2005}.  \citet{Feng_2021} note that secondary's HR position is decently inconsistent with its mass and the age implied by the evolved primary.  Our inference captures this inconsistency as a sub-solar mixing length, allowing the star to be cooler than standard models would predict.  \citet{Pawar_2024} provided an updated eclipse solution which alleviated some of the discrepancy but we still find that our variable $\amlt$ fit offers a far better solution than standard models.

Lastly is V432 Aur, a metal-poor ($\FeH = -0.6$) binary containing a subgiant (component B) and main-sequence star (component A).  If the subgiant has remained tidally synchronized then its orbital period implies it is rapidly rotating $(\Ro \sim 0.09)$, yet we infer $\amltB \approx \amltA$ in contrast to the picture of magnetic inflation in other rapid rotators.  The temperature of the main-sequence component also appears anomalously hot relative to standard $\solaramlt$ models and so this system would benefit from a more focused re-analysis where it might prove a stringent test to metal-poor stellar models.

\section{Non-solar Mixing Lengths in Interferometric Stars}
\label{section:irad_results}
The $\solaramlt$ fits to the \interferometric\ stars must be treated differently due to the lack of a mass constraint.  In these fits we have $\boldsymbol D = \{R, ~\Teff, ~\FeHs \}$, $\boldsymbol \theta = \{\mathrm{EEP}, ~\Mi, ~\FeHi \}$, and $\nu = 0$; the model is saturated.  So long as $\boldsymbol D$ does not fall egregiously outside of stellar parameter space then the fit can always find a model that is consistent with the observations, and without any further constraints we cannot form a meaningful $\chi^2$ to judge the quality of that solution.  With no a priori constraint on the mass and with $\FeHi$ well constrained by $\FeHs$ the only remaining way to assess the quality of the solution is to look for physical inconsistencies in the inferred age.

\begin{figure*}[t!]
\epsscale{1.0}
\plotone{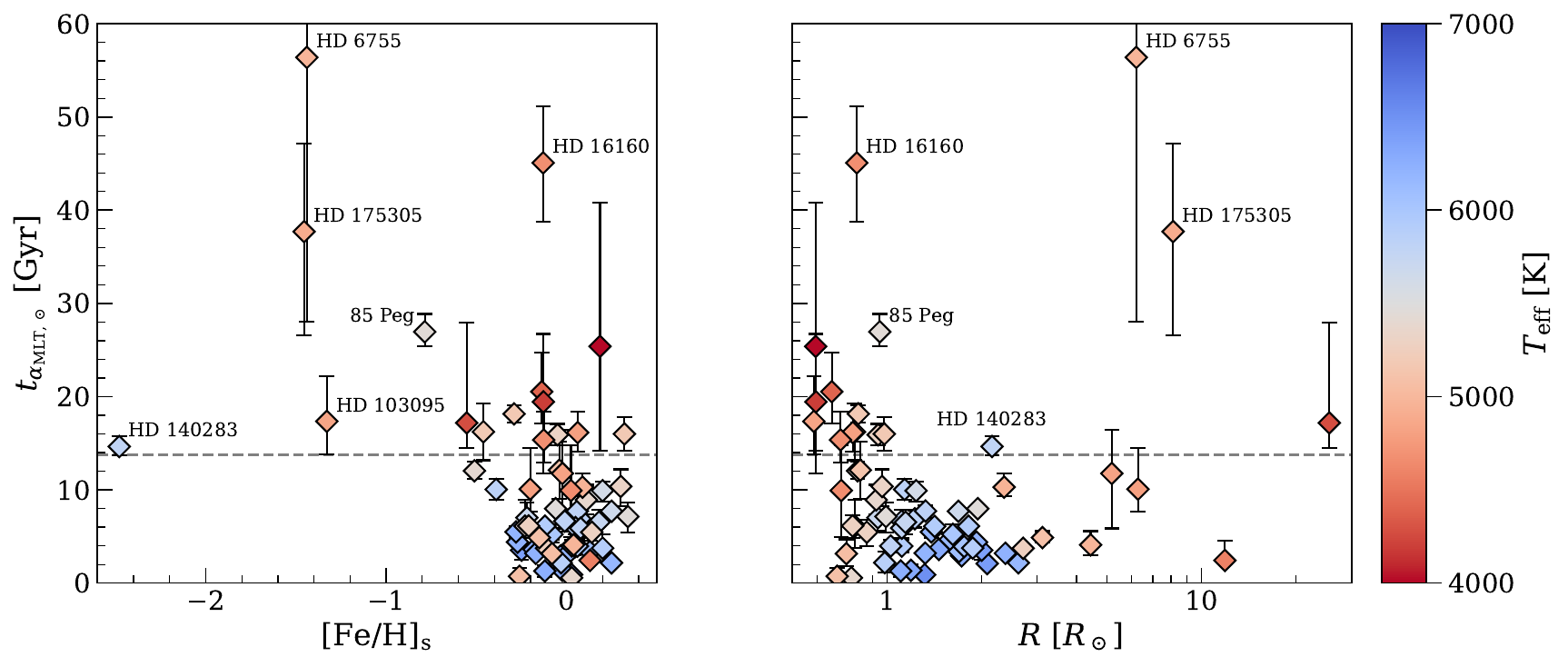}
\caption{Ages inferred for the \interferometric\ stars when assuming $\amlt = \solaramlt$, plotted against surface metallicity and radius.  We find two sets of stars with unphysically old ages \citep[$t~ > 13.8$ Gyr;][]{PlanckCollaboration_2020}: cool dwarfs and metal-poor stars.  The masses and initial metallicities of the cool dwarfs are well constrained by their luminosities and surface metallicities, The majority of solar metallicity stars with $\Teff \lesssim 4500$ K have ages $>8$ Gyr, in contrast with their hotter counterparts and inconsistent with Galactic chemical evolution.  Metal poor stars are expected to be old \citep[$8-12$ Gyr;][]{Haywood_2018, Bonaca_2020, Xiang_2022}, but the ages we infer are consistently too old, with some several times older than the age of the universe.  Standard models assuming $\amlt = \solaramlt$ fail to reproduce these stars' observed radii at realistic ages.\label{figure:irad_solaramlt_ages}}
\end{figure*}

In Figure \ref{figure:irad_solaramlt_ages} we plot the $\solaramlt$ inferred ages versus observed $\FeHs$ and radius.  We find a surplus of stars at unphysically old ages ($t~ > 13.8$ Gyr) and so despite the fits having zero degrees of freedom, the \interferometric\ sample is still informative and capable of revealing the presence of some insufficiency in the $\solaramlt$ models.  We can separate the stars with unphysically old ages into two groups: cool dwarfs and metal-poor stars.  

The cool dwarfs are relatively metal-rich ($\FeHs \gtrsim -0.5$) main-sequence stars with $\Teff \lesssim 5400$ K.  They are all nearby (they can be resolved interferometrically) and are all on orbits typical of stars in the Galactic thin disk, so by every indication they are ordinary lower main-sequence dwarfs.  We expect such stars in the solar neighborhood to be relatively young \citep[$t~ \lesssim 8$ Gyr;][]{Nissen_2015}, so even if we consider that the absolute age scale of our stellar models is bound to be slightly imperfect we would still not expect such anomalous ages.  The mass inferred for these stars is to first order constrained by their luminosity, and since these stars are largely convective the effects of atomic diffusion are minimal ($\FeHi \approx \FeHs$), leaving EEP, i.e., age as the sole parameter left to vary in order to match the observed radius and temperature.  The unphysically old age can therefore be interpreted as the model trying to reproduce a radius (temperature) that is too seemingly large (cool) for a star of that mass, again indicative of some form of radius inflation.

\begin{figure}[t!]
\epsscale{1.0}
\plotone{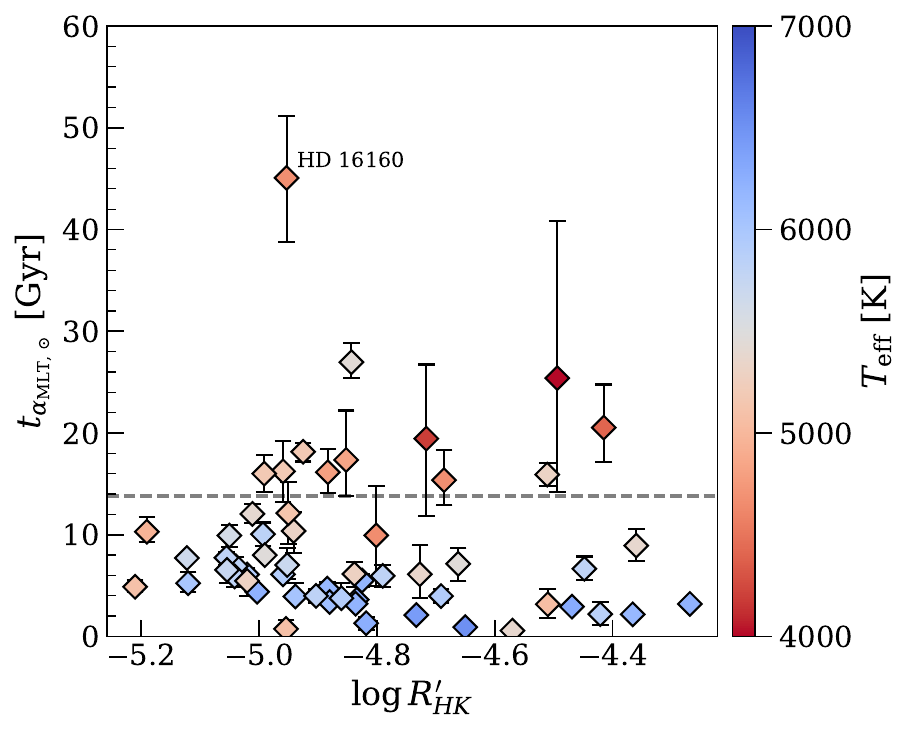}
\caption{$\solaramlt$ implied ages versus Ca II H\&K emission $\log(R'_{HK})$ for the \interferometric\ stars.  Higher levels of chromospheric activity are present in about half of the coolest dwarfs, pointing towards magnetic inflation as an explanation for their unphysically old $\solaramlt$ ages.  However, there remains several cool dwarfs at unrealistic ages that have low levels of activity and so their seeming inflation remains unexplained.\label{figure:irad_solaramlt_ages_vs_logRHK}}
\end{figure}

We compiled rotation periods and activity indicators for the \interferometric\ stars to see if the results for these cool dwarfs are consistent with magnetic inflation as we saw in the \debs\ (Section \ref{subsection:fast_rotators}).  Our compilation can be found in Table \ref{table:irad_activity_indicators} and Figure \ref{figure:irad_solaramlt_ages_vs_logRHK} plots the $\log(R'_{HK})$ chromospheric emission index versus the $\solaramlt$ inferred age.  While some of the unphysically old, cool dwarfs do indeed have high levels of activity it is not ubiquitous, and the inflation in about half of the group remains unexplained.

The second group of stars with unphysical $\solaramlt$ ages are the metal-poor stars, and in fact all of the stars with $\FeHs \lesssim -0.6$ in our \interferometric\ sample have median ages $> 13.8$ Gyr.  Metal poor stars are expected to be old, having formed before significant chemical evolution had time to occur, and so finding ages slightly older than the age of the universe is not unsurprising when considering again that the absolute age scale of our models is likely imperfect.  An imperfect age scale is still a flaw in the models that can be improved; accurate absolute stellar ages are a challenging but worthy goal to strive towards.  But regardless of the absolute age scale there are cases among the metal-poor stars of ages dramatically inconsistent with the age of the universe, with HD 175305, HD 6755, and 85 Peg in particular having median ages $t_{\solaramlt} > 25$ Gyr, marking a far more egregious issue with standard stellar models at low metallicity. 

The variable $\amlt$ fits to the \interferometric\ stars must again be interpreted differently than those of the \debs, except this time the inclusion of $\amlt$ as a free parameter makes the fit overparameterized.  The posteriors $p(\boldsymbol \theta | \boldsymbol D)$ are therefore highly degenerate and can never infer a ``single $\amlt$", instead describing the hypersurface of models that are consistent with the observations.  

\begin{figure}[t!]
\epsscale{1.0}
\plotone{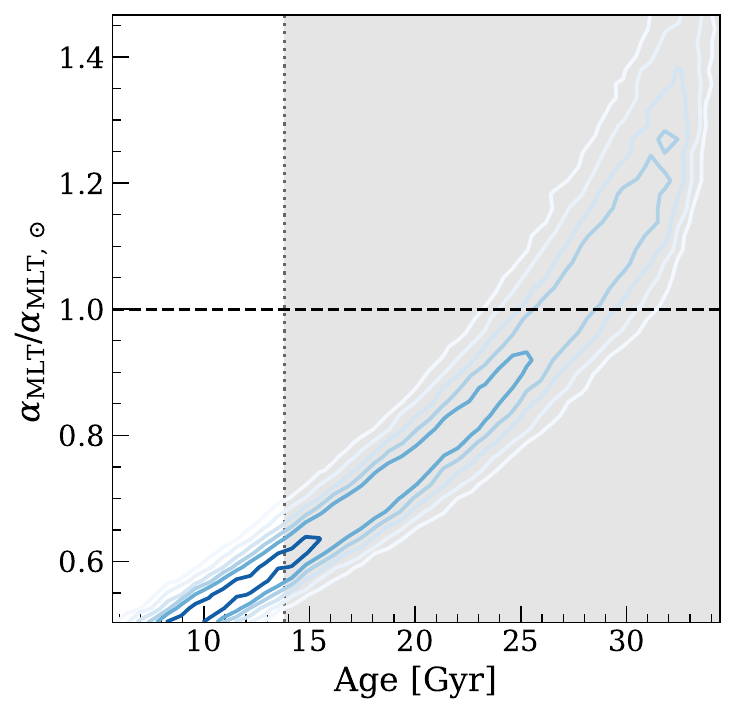}
\caption{Age$-\amlt$ posterior for the metal-poor dwarf 85 Peg, part of the \interferometric\ sample.  The lack of a mass constraint leaves the posterior highly degenerate, but the radius, $\Teff$, and $\FeHs$ still constrain the slope of the posterior such that only models with a sub-solar mixing length are consistent with observations at realistic ages.
\label{figure:85Peg_age_amlt_posterior}}
\end{figure}

Information can still be gleaned from these highly degenerate posteriors, which we demonstrate in Figure \ref{figure:85Peg_age_amlt_posterior} with the metal-poor ($\FeHs = -0.79$) dwarf 85 Peg.  The degeneracy in $t~ - \amlt$ space ensures that there is no single $\amlt$ estimate, but the slope of the posterior is such that only models with $\amlt < \solaramlt$ are consistent with measurements at physically realistic ages.  This is the result of Figure \ref{figure:irad_solaramlt_ages} re-framed; models either need a sub-solar mixing length or need to evolve to an unphysically old age in order to reach the observed radius at the star's luminosity and metallicity.  

\begin{figure}[t!]
\epsscale{1.0}
\plotone{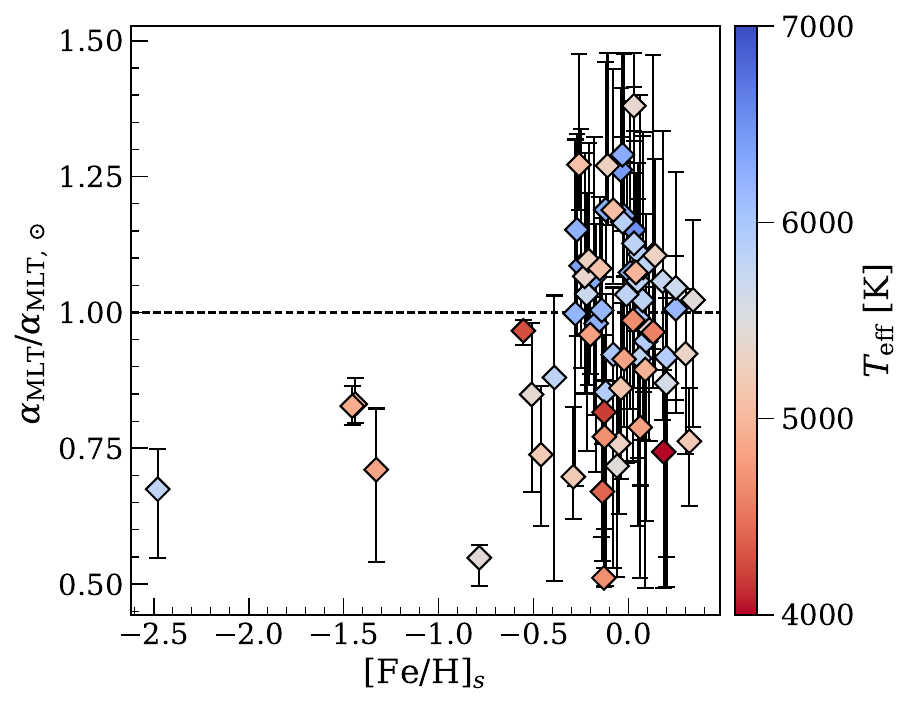}
\caption{Bounds on the mixing length inferred in the \interferometric\ stars as a function of metallicity when $t~ < 13.8$ Gyr is enforced.  Models require a subsolar mixing length in order to match the radii and temperatures of low metallicity stars at physically reasonable ages.  We find $\amlt / \solaramlt \approx 0.5 - 0.75$ at metallicities $\FeHs \lesssim -0.5$.  
\label{figure:irad_feh_amlt}}
\end{figure}

This age-implied sub-solar mixing length can be seen as a generic feature in all of the metal-poor \interferometric\ stars and was also seen in $\mu$ Cas A (Section \ref{subsection:discuss_muCas}).  Figure \ref{figure:irad_feh_amlt} shows the \interferometric\ sample's mixing lengths as a function of their metallicities when we restrict to only the posterior samples with $t~ < 13.8$ Gyr.  The marked point is the median of the marginal age posteriors and the error bars span the 16th$-$84th percentile interval of $\amlt$ for these age restricted samples.  Stars with $\FeH \approx 0.0$ scatter around $\solaramlt$ but at $\FeH \lesssim -0.5$ their mixing lengths clearly tend sub-solar.

\begin{figure}[t]
\epsscale{1.0}
\plotone{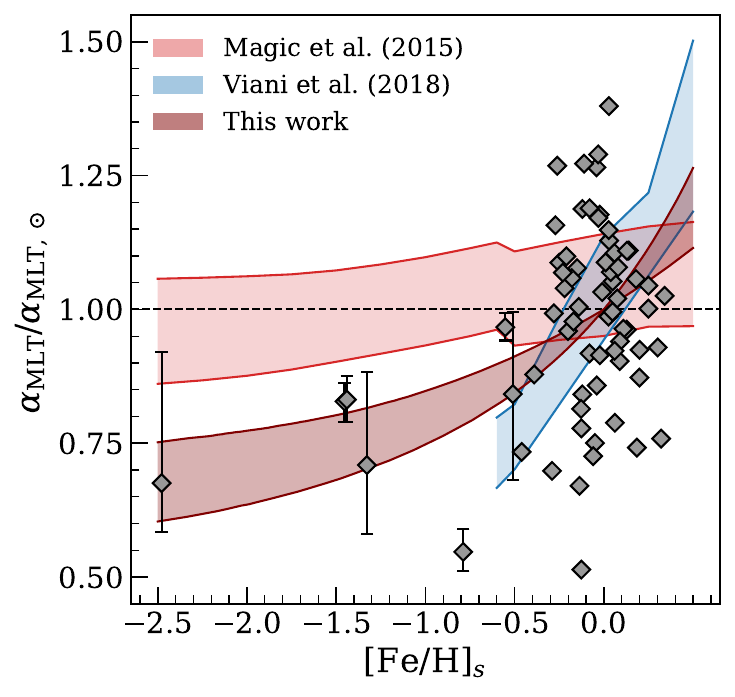}
\caption{Mixing length as a function of metallicity for the interferometric stars, compared to the metallicity dependence predicted by the 3D STAGGER simulation \citep{Magic_2015} and that observed in asteroseismic samples \citep{Viani_2018}.  Simulations do not predict significant changes to mixing length at low metallicity, inconsistent with our findings.  We sketch a simple extension to the \citet{Viani_2018} calibration that roughly captures the sub-solar mixing lengths we infer for the most metal-poor stars.\label{figure:irad_feh_amlt_sketch}}
\end{figure}

This result is consistent with previous empirical studies of mixing length in metal-poor stars.  In particular \citet{Bonaca_2012, Tayar_2017, Viani_2018, Li_2024} each observe a linear correlation of decent strength ($\amlt/\solaramlt \propto 0.1 - 0.6 \times \FeHs$) between mixing length and metallicity, though the metallicity range of their samples were smaller than ours and centered around solar metallicity.  Our results with the \interferometric\ sample are too noisy to confirm or deny a linear correlation about $\FeH \approx 0.0$ but our sample does show that the correlation with metallicity seemingly continues and flattens out at lower metallicity, saturating at a value of $\amlt / \solaramlt \approx 0.50 - 0.75$ for stars beyond $\FeHs \lesssim -1.0$, in very good agreement with \citet{Joyce_2018}.  We note that a flattening of any $\FeH - \amlt/\solaramlt$ correlation at low metallicities is expected and necessary because $\FeH$ is a relative logarithmic quantity and so the absolute metallicity $Z$ is converging to zero.

In Figure \ref{figure:irad_feh_amlt_sketch} we compare the inferred mixing lengths for the \interferometric\ stars against the metallicity dependence of the \citet{Magic_2015} and \citet{Viani_2018} mixing length calibrations.  Both calibrations also predict $\amlt$ to vary with $\Teff$ and $\logg$ and so at each metallicity we marginalize over the ($\Teff$, $\logg$) values of a 5 and 10 Gyr isochrone. 

The magnitude of variation in $\amlt$ with metallicity that we observe in the \interferometric\ sample is far larger than predicted by 3D RHD simulations.  The STAGGER grid \citep{Magic_2013, Magic_2015} is the only suite of simulations to test metallicities as low as $\FeH < -1.0$, reporting a metallicity dependence of at most only $\amlt/\solaramlt \propto -0.02 \times \FeH$ (there is some ambiguity in how to define the ``mixing length" in a simulation).  The sign of the metallicity dependence predicted by the simulations is also in conflict with empirical results.  The near surface superadiabatic entropy gradient in 1D MLT models is generically inconsistent with the mean profile predicted in simulations, and this inconsistency only worsens as metallicity decreases \citep[e.g., Figure 1 of][]{Magic_2015}.  The stark contrast in metallicity dependence could be in part due to the difference in shape of this superadiabatic jump, and new 1D models with an improved near surface entropy gradient might lead to a resolution \citep{Magic_2016, Spada_2018, Spada_2019, Sonoi_2019, Spada_2021, Manchon_2024}.  \citet{Li_2024} also suggest that the peculiar helium to metal enrichment ratio ($\dydz$) assumed in the STAGGER grid makes direct comparison to empirical mixing length calibrations difficult.  

Our \interferometric\ sample extends two dex lower in metallicity than the \citet{Viani_2018} sample and so in Figure \ref{figure:irad_feh_amlt_sketch} we sketch a simple extension to the $\amlt-$metallicity dependence they report.  This toy model is simply an exponential as a function of $\FeH$ that 1) asymptotes at sub-solar $\amlt/\solaramlt$ at the lowest metallicities, 2) matches the \citet{Viani_2018} slope at $\FeH = 0$, and 3) ensures $\amlt/\solaramlt = 1$ at $\FeH = 0$.    This functional form can be translated to give mixing length as a power law on the total metal to hydrogen mass fraction, i.e., $\amlt/\solaramlt \propto (Z/X)^b$.  We emphasize that this functional form is completely arbitrary and is only meant as a toy model for how mixing length might vary at the lowest metallicities.  More work is needed and crucially more precise stellar radii and temperatures are needed at low metallicities in order to make a proper calibration.

\section{Mixing Length in Cluster \debs}
\label{section:cluster_results}
\begin{figure*}[t]
\epsscale{1.0}
\plotone{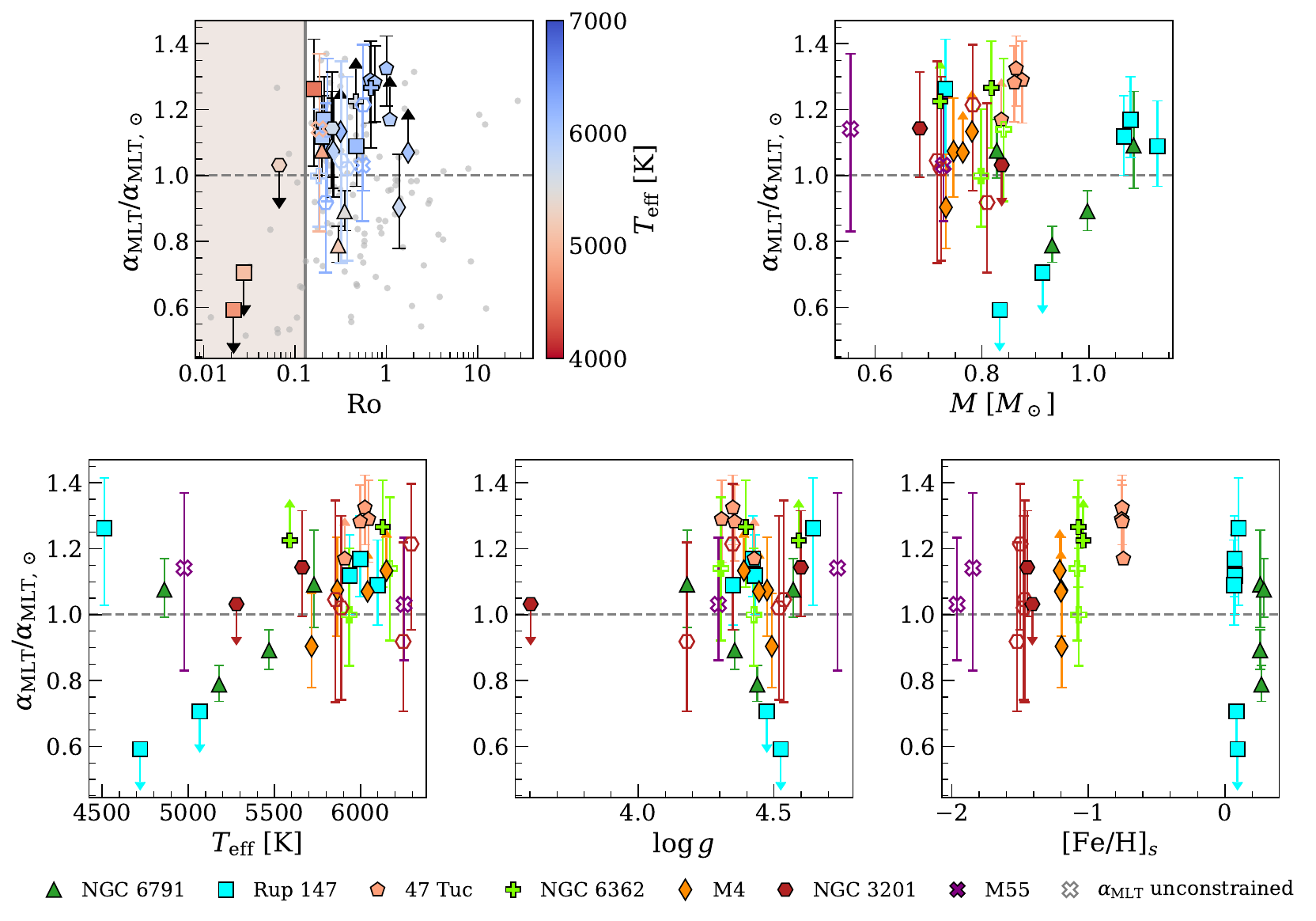}
\caption{Mixing length $\amlt/\solaramlt$ versus stellar parameters for the cluster DEBs (Table \ref{table:clusters}).  Like the field DEBs (Section \ref{section:debs_results} we observe inflation among the cooler rapid rotators but do not find any other correlations with stellar parameters.  The relative mixing lengths within a cluster do not indicate coherent variation with stellar parameters.  In particular the globular cluster DEBs do not show a metallicity dependent mixing length despite being extremely metal-poor \citep[Section \ref{section:irad_results};][]{Tayar_2017, Viani_2018, Joyce_2018, Li_2024}, in fact the GC DEBs tend towards $\amlt > \solaramlt$.  It is possible that non-standard helium abundances are at play in these cases (see text).\label{figure:cluster_amlt_summary}}
\end{figure*}

Our isochrone fits to the clusters require all components of all member binaries (Table \ref{table:clusters}) to share a single age and initial composition.  This makes the clusters sensitive tests of relative mixing length, and particularly sensitive to any potential mass ordering.  Figure \ref{figure:cluster_amlt_summary} summarizes our findings and plots the $\amlt$ of the cluster DEBs against stellar parameters.  Each cluster is given its own color and symbol unless otherwise specified.  In the $M$, $\Teff$, $\logg$, and $\FeHs$ panels we retain stars with uninformative $\amlt$ posteriors as defined in Section \ref{section:debs_results}, plotting them with open symbols.  

The first panel plots the Rossby number of cluster DEBs (assuming tidal synchronization) versus $\amlt$, in analogy with Figure \ref{figure:debs_amlt_vs_Ro}.  Again we find sub-solar mixing lengths for cool, rapidly rotating stars, consistent with the picture of magnetic inflation (Section \ref{subsection:fast_rotators}).  Beyond this we find no correlation between $\amlt$ and mass, $\Teff$, $\logg$, or $\FeHs$, matching our findings with the field DEBs (Section \ref{section:debs_results}).  

We again find no convincing evidence for a mass ordering among the mixing lengths, whether that is within individual binaries or across an entire cluster.  There is no coherent gradient of mixing length as a function of either $\Teff$ or $\logg$.  For example, $\amlt$ is well constrained in the three lowest mass stars of NGC 6791 and yet they clearly do not indicate systematic variation with Kiel position, instead seeming to just scatter near $\solaramlt$.  

Despite the two dex range in metallicity the cluster DEBs do not appear consistent with a metallicity dependent mixing length like we see in the interferometric stars (Section \ref{section:irad_results}).  In fact the globular cluster DEBs tend towards super-solar mixing lengths, in contrast with other metal-poor mixing length calibrations \citep{Joyce_2018}.  A super-solar mixing length indicates that these stars are actually smaller than standard models predict.  Several of the globular cluster DEBs have $\amlt$ posteriors close to or fully up against the prior edge, indicating that the amount of radius deviation is rather large.  

This is another situation where a non-standard prescription for different model parameter, namely helium abundance, may be more appropriate in order to explain model discrepancies.  Our MIST models assign helium abundance as a function of metallicity, $Y_i = Y_p + (\frac{\Delta Y}{\Delta Z})Z_i$, and so at low metallicity our models assume a nearly primordial helium abundance.  However, many globular clusters are observed to have multiple populations that vary in their helium abundances \citep{Bragaglia_2010, Conroy_2010, Milone_2011}.  An enhanced helium abundance increases the mean molecular weight of the star, shrinking it relative to a star of the same mass and age with a nearly primordial helium abundance as in our models.  An enhanced helium could therefore explain why our models are larger than the observed radii of many of the GC DEBs, and explain why we infer such large values for $\amlt$.  The globular cluster DEBs likely require a more thorough investigation that simultaneously varies both $Y$ and $\amlt$ beyond their solar-calibrated values \citep{Jiaqi_2023}.

\section{Caveats \& Limitations}
\label{section:caveats}
In this section we briefly identify some caveats and additional sources of uncertainty that impact our analysis, and we discuss potential routes to mitigate such uncertainties in the future.  

Our \debs\ sample is a heterogeneous compilation from the literature with the methods used to characterize each system varying from author to author.  This heterogeneity is bound to add additional noise into any global $\amlt$ trends we attempt to test with the \debs\ sample.  An effort to improve and homogenize the $\Teff$ and $\FeH$ measurements in particular would be useful as these quantities are less precise than the masses or radii, and methods of spectral analysis for these SB2 eclipsing systems are varied.  

We intentionally selected binary stars that should most likely represent single star evolution, but it is possible that some of our binaries are the products of mass transfer and mergers despite now appearing as detached systems.  The rate of close companions and triples each increases with mass \citep{Moe_2017} so this additional source of noise will be more prevalent among our higher mass and our evolved binaries. 

The neural network emulation is imperfect and contributes an additional numerical uncertainty to the mapping between observables and inferred parameters like $\amlt$.  Our tests indicate that this source of uncertainty is small compared to the measurement uncertainties in most cases, but sharp evolutionary features like the convective hook are difficult to emulate and the numerical uncertainty around these phases will be worse (Figure \ref{figure:nn_validation}).

In this work we explored models with a non-solar mixing length but it is perfectly plausible that some other uncertainty or approximation in 1D models is the culprit behind the poor $\solaramlt$ fits of some systems.  Chief among these other confounding variables are the alpha-element abundance $\aFe$, the helium abundance $Y$, the strength of atomic diffusion, and the core overshooting parameter $\fovcore$ (see Sections \ref{subsection:discuss_Procyon}, \ref{section:cluster_results}).  

\begin{figure}[t!]
\epsscale{1.0}
\plotone{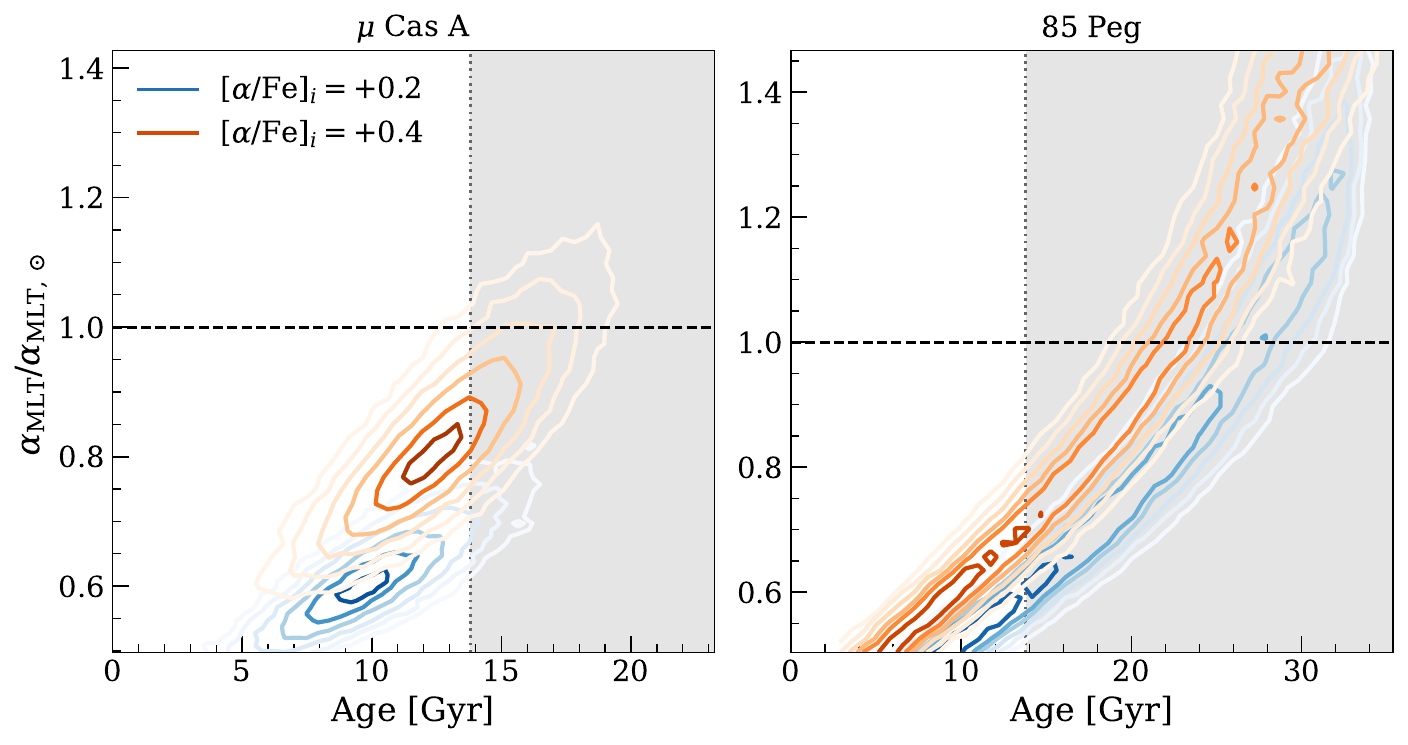}
\caption{Sensitivity of our inferred mixing lengths to initial alpha element enhancement $\aFei$.  Underestimating the star's true $\aFe$ leads to a smaller value of $\amlt$.  This can be understood as lowering the total metal to hydrogen fraction [M/H] at fixed iron metallicity [Fe/H], leading to a lower $\amlt$ via the metallicity$-\amlt$ correlation sketched in Figure \ref{figure:irad_feh_amlt_sketch}.  The majority of DEBs do not have detailed abundance measurements and so an unknown $\aFe$ remains an unavoidable source of uncertainty.  $\mu$ Cas A and 85 Peg are both measured to have $\aFes \approx +0.3$ \citep{Casamiquela_2026}.\label{figure:afe_sensitivity}}
\end{figure}

The vast majority of \debs\ do not have chemical abundances reported for elements beyond iron and so we simply cannot account for variations beyond solar-scaled abundances, thus we chose to crudely fix $\aFe$ as a function of $\FeH$ when we built our model grids.  In Figure \ref{figure:afe_sensitivity} we use two stars that do have detailed abundances available, $\mu$ Cas A and 85 Peg, to assess the effect of an incorrect $\aFei$ on our inferred mixing lengths.  An underestimation of $\aFei$ by 0.2 dex tends to decrease $\alpha/\solaramlt$ by $\approx 0.2$ in these metal-poor dwarfs, a quite sizable systematic shift.  This is undoubtedly a major additional source of scatter in the inferred $\amlt$'s of the \debs\ sample that is currently unavoidable, further encouraging a homogeneous spectroscopic analysis.

Our interferometric stars were selected from the GBSv3 catalog \citep{Soubiran_2024} and therefore do have homogeneous measurements of spectroscopic parameters and as well as detailed abundances, thus a more careful accounting of the stars' true $\aFe$ is possible.  However, we argue that our results for the interferometric stars in particular are still robust to an imperfect $\aFe$, in particular the mixing length$-$metallicity correlation we observe in Section \ref{section:irad_results}.  The interferometric stars' posteriors are already highly degenerate due to the lack of a mass constraint, and Figure \ref{figure:afe_sensitivity} demonstrates that an incorrect $\aFei$ mostly shifts the posterior and does not significantly alter its overall slope.  It is the posterior slopes combined with the requirement of a realistic age that underpins our finding that $\amlt < \solaramlt$ at low metallicity and so we do not expect $\aFe$ to change the qualitative result we find. 

An alternative way to view this is that $\aFe$ sets the mapping between the iron metallicity [Fe/H] and the total metals to hydrogen ratio [M/H], so an underestimated $\aFei$ leads to an underestimated [M/H] at the same value of [Fe/H].  If mixing length is sensitive to metallicity it will be to the total metal content of the star's atmosphere and not just the iron abundance, and so we would expect the underestimated [M/H] to result in a lower $\amlt$ as well.  The incorrect $\aFe$ amounts to a horizontal shift along the metallicity$-\amlt$ correlation we sketch in Figure \ref{figure:irad_feh_amlt_sketch}, and the resulting shift in $\amlt$ is consistent with the effect observed in Figure \ref{figure:afe_sensitivity}.

We considered the masses and radii of each component in a DEB as independent observables, but this is technically an overly ideal assumption.  In reality these quantities are correlated through the modeling of the light curve and radial velocities, but unfortunately these correlations are often not reported.  These correlations would propagate through to our $\amlt$ posteriors and could potentially alter our results, in particular cases where the two components in the binary display an ordering in relative mixing length (Section \ref{subsection:debs_results_relative_amlt}).  We encourage the reporting of the full covariance results from the light and radial velocity curve analyses as properly incorporating them can magnify the constraining power of eclipsing binaries on stellar models \citep{Maxted_2020, Miller_2020}.

\section{Conclusions}
\label{section:conclusion}
In this work we used the precise geometric radii of detached eclipsing binaries (DEBs) and interferometrically resolved stars to test standard \misttwo\ models that use a solar-calibrated mixing length parameter ($\solaramlt$) against models with a variable mixing length ($\alpha$).  In Table \ref{table:debs_results} we flag systems that are inconsistent with $\solaramlt$ models and those whose variable $\amlt$ fits strongly indicate that a non-solar mixing length is capable of better explaining observations.  

We found that approximately one third of the DEB systems are poorly fit by $\solaramlt$ models, a fraction far larger than expected by statistical chance.  Several of these systems prove to be cool ($\Teff < 5400$ K), rapidly rotating ($\Ro < 0.13$) stars with radii (temperatures) too large (cool) than standard $\solaramlt$ model predict.  These are examples of the well known effect of rotationally driven magnetic inflation of convective stars \citep{Lopez-Morales_2007}, and our isochrone fits with variable $\amlt$ models capture this as a sub-solar mixing length.  We also find cases where hotter or more slowly rotating stars appear similarly inflated, though it is unclear whether the same magnetic inflation mechanism is operating in these stars given the different properties in their convective envelopes.

Beyond this rotationally induced magnetic inflation we find no population level correlations between stellar parameters and the inferred mixing length of \debs.  This is in contrast to asteroseismic samples wherein mixing length appears to vary strongly with stellar parameters, particularly metallicity \citep{Bonaca_2012, Metcalfe_2014, Creevey_2017, Tayar_2017, Viani_2018, Li_2024}.  It is unclear if this is simply due noise or systematics in the \debs\ measurements or if eclipsing binaries behave qualitatively different as a population, the possibility of which calls into question their validity as calibrators of single star models.  We highlight that homogeneous spectral analyses including detailed abundances measurements would greatly refine the diagnostic power of the DEBs.

We identify several particular binaries that cannot be fit by a single $\solaramlt$ isochrone and that also display evidence for variation in the mixing length (Section \ref{subsection:case_studies}).  It is possible that some of these cases could be better explained by other changes to the models e.g., non-standard helium abundance or core overshooting.  Nevertheless these systems are ideal candidates for future case studies into mixing length and can serve to better calibrate stellar models beyond the Sun.

We highly recommend the use of the interferometric sample and Gaia Benchmark star catalog \citep{Blanco-Cuaresma_2014, Soubiran_2024, Casamiquela_2026} more broadly as a means to test stellar models and non-solar mixing length prescriptions.  We have demonstrated that the combination of radius measurements and the catalog's careful and homogeneous temperature and metallicities makes it discerning.  A campaign to measure the rotation periods and activity levels in all of these stars would help to avoid confounding signals from magnetic inflation as is seen in the \debs, improving the diagnostic power of this sample at cool temperatures.

The interferometric stars reveal a clear issue where standard models underpredict the radii of metal-poor stars.  We find that this issue can be neatly explained by decreasing mixing length at low metallicity.  While we cannot make a meaningful quantitative calibration, this result is in great qualitative agreement with other empirical work on the mixing length \citep{Bonaca_2012, Metcalfe_2014, Creevey_2017, Tayar_2017, Viani_2018, Joyce_2018, Li_2024}.  3D RHD simulations fail to predict the variation in mixing length with metallicity that we observe and so we caution the use of simulation based mixing length calibrations \citep[i.e., entropy calibrations][]{Magic_2015, Magic_2016, Tanner_2016, Spada_2018, Spada_2019, Sonoi_2019, Spada_2021, Manchon_2024} until this discrepancy is explained.  

Mixing Length Theory is only an approximate description of convection in 1D and so the exact physical mechanism causing the models to underpredict metal-poor radii remains an open question.  The answer could lie in some facet of convection that MLT fails to capture, but given the seeming insensitivity of 3D simulations to metallicity this facet is be difficult to identify.  The culprit could instead be inaccuracies in input physics (e.g., opacities) or other internal mixing processes (e.g., atomic diffusion), and we intend to explore these possibilities in future work. 

Precise, classical, absolute stellar parameters remain an effective constraint on the stellar models and stellar ages that underpin our understanding of the evolution of exoplanetary and stellar populations in galaxies including and beyond our own.  They reveal that standard assumptions such as solar-calibrated parameters are insufficient if we wish to achieve absolute stellar ages that are accurate to $\lesssim 10 \%$.  The classical calibrators we identify in this work, in combination with asteroseismic calibrators, will enable a more sophisticated future treatment of systematic uncertainties in stellar models, wherein troublesome parameters like mixing length, core overshooting, helium abundance, diffusion etc. can be simultaneously constrained and their uncertainties self-consistently propagated into the age inferences of ordinary field stars.  

\appendix
\begin{deluxetable*}{lccc@{\hspace{2.5em}}lccc@{\hspace{2.5em}}lccc}
\tablecaption{Rotation periods and Ca\,{\sc ii} activity indicators of the Interferometric Radius Stars.\label{table:irad_activity_indicators}}
\tablehead{\colhead{Star} & \colhead{$P_{\rm rot}$ (d)} & \colhead{$S$} & \colhead{$\log R'_{\rm HK}$} & \colhead{Star} & \colhead{$P_{\rm rot}$ (d)} & \colhead{$S$} & \colhead{$\log R'_{\rm HK}$} & \colhead{Star} & \colhead{$P_{\rm rot}$ (d)} & \colhead{$S$} & \colhead{$\log R'_{\rm HK}$}}
\startdata
10 Tau & \nodata & 0.139$^{4}$ & -4.86$^{2}$ & $\beta$ Com & 12.3$^{8}$ & 0.198$^{2}$ & -4.69$^{2}$ & HD 158633 & \nodata & 0.174$^{4}$ & -4.96$^{2}$ \\
11 LMi & 18.0$^{1}$ & 0.274$^{2}$ & -4.66$^{2}$ & $\beta$ Hyi & \nodata & 0.158$^{2}$ & -4.91$^{2}$ & HD 167042 & \nodata & 0.130$^{4}$ & \nodata \\
12 Oph & 21.1$^{9}$ & 0.330$^{2}$ & -4.57$^{2}$ & $\beta$ Vir & \nodata & 0.160$^{2}$ & -4.85$^{2}$ & HD173701 & \nodata & 0.190$^{4}$ & -4.98$^{2}$ \\
15 Sge & \nodata & 0.199$^{2}$ & -4.72$^{2}$ & $\chi^1$ Ori & 5.2$^{13}$ & 0.311$^{2}$ & -4.42$^{2}$ & HD 219134 & 42.3$^{7}$ & 0.292$^{4}$ & -4.86$^{2}$ \\
16 Cyg A & 23.8 $\pm$ 1.8$^{6}$ & 0.150$^{4}$ & -4.96$^{2}$ & $\chi$ Cnc & \nodata & 0.164$^{2}$ & -4.79$^{2}$ & $\iota$ Per & \nodata & 0.153$^{2}$ & -4.92$^{2}$ \\
16 Cyg B & 23.2 $\pm$ 3.2$^{6}$ & 0.154$^{2}$ & -4.97$^{2}$ & $\delta$ Eri & \nodata & 0.136$^{4}$ & -5.19$^{5}$ & $\iota$ Psc & \nodata & 0.156$^{2}$ & -4.84$^{2}$ \\
18 Sco & 22.7 $\pm$ 0.5$^{10}$ & 0.172$^{2}$ & -4.87$^{2}$ & $\epsilon$ Eri & 11.1 $\pm$ 0.1$^{14}$ & 0.447$^{2}$ & -4.51$^{2}$ & $\xi$ Peg & \nodata & 0.142$^{2}$ & -4.94$^{2}$ \\
20 LMi & \nodata & 0.147$^{4}$ & -4.98$^{2}$ & $\epsilon$ Ind & \nodata & 0.668$^{2}$ & -4.52$^{2}$ & $\lambda$ Aur & \nodata & 0.146$^{4}$ & -4.83$^{2}$ \\
24 Sex & \nodata & 0.128$^{4}$ & \nodata & $\eta$ Boo & \nodata & 0.319$^{2}$ & -4.37$^{2}$ & $\mu$ Her & \nodata & 0.151$^{4}$ & -5.13$^{2}$ \\
36 UMa & \nodata & 0.173$^{2}$ & -4.78$^{2}$ & $\eta$ Ser & \nodata & 0.131$^{4}$ & \nodata & $o^2$ Eri & \nodata & 0.199$^{2}$ & -4.92$^{2}$ \\
51 Peg & \nodata & 0.149$^{2}$ & -5.04$^{2}$ & $\gamma$ Ser & \nodata & 0.156$^{2}$ & -4.85$^{2}$ & $\pi^3$ Ori & \nodata & 0.214$^{2}$ & -4.57$^{2}$ \\
54 Psc & \nodata & 0.173$^{4}$ & -5.12$^{2}$ & HD 166 & \nodata & 0.478$^{2}$ & -4.33$^{2}$ & Procyon & \nodata & 0.193$^{4}$ & -4.11$^{2}$ \\
61 Cyg A & 35.4 $\pm$ 1.9$^{15}$ & 1.284$^{2}$ & -4.23$^{2}$ & HD 4628 & 38.0$^{18}$ & 0.230$^{2}$ & -4.91$^{2}$ & $\rho^1$ Cnc & 39.0$^{3}$ & 0.179$^{4}$ & -4.12$^{2}$ \\
61 Cyg B & 48.0 $\pm$ 0.7$^{17}$ & 0.975$^{2}$ & -4.72$^{2}$ & HD 4747 & \nodata & 0.246$^{4}$ & -4.73$^{2}$ & $\sigma$ Dra & 27.7 $\pm$ 0.8$^{9}$ & 0.221$^{2}$ & -4.81$^{2}$ \\
61 UMa & 17.1$^{9}$ & 0.309$^{2}$ & -4.54$^{2}$ & HD 5015 & \nodata & 0.146$^{2}$ & -4.94$^{2}$ & $\tau$ Cet & \nodata & 0.152$^{2}$ & -5.04$^{2}$ \\
70 Vir & \nodata & 0.165$^{2}$ & -4.96$^{2}$ & HD 16160 & 48.0$^{9}$ & 0.228$^{4}$ & -4.95$^{2}$ & $\theta$ Boo & \nodata & 0.255$^{2}$ & -4.47$^{2}$ \\
85 Peg & \nodata & 0.184$^{2}$ & -4.84$^{2}$ & HD 38858 & \nodata & 0.125$^{2}$ & -5.18$^{2}$ & $\theta$ Per & \nodata & 0.176$^{4}$ & -4.70$^{2}$ \\
94 Cet & \nodata & 0.160$^{2}$ & -4.86$^{2}$ & HD 79210 & \nodata & 1.786$^{2}$ & -4.49$^{2}$ & $\theta$ UMa & \nodata & 0.336$^{2}$ & -4.27$^{2}$ \\
107 Psc & 35.0$^{9}$ & 0.193$^{4}$ & -4.97$^{2}$ & HD 103095 & \nodata & 0.200$^{2}$ & -4.85$^{2}$ & $w$ Her & \nodata & 0.158$^{4}$ & -4.92$^{2}$ \\
110 Her & \nodata & 0.202$^{4}$ & -4.61$^{2}$ & HD 130948 & 7.8$^{12}$ & 0.285$^{2}$ & -4.45$^{2}$ &  &  &  &  \\
$\beta$ Aql & \nodata & 0.135$^{4}$ & -5.21$^{16}$ & HD 131977 & 44.6$^{11}$ & 0.380$^{2}$ & -4.87$^{2}$ &  &  &  &  \\
\enddata
\tablerefs{(1) \citet{Radick_1987}; (2) \citet{Saikia_2018}; (3) \citet{Bourrier_2018}; (4) \citet{Isaacson_2010}; (5) \citet{Marsden_2014}; (6) \citet{Davies_2015}; (7) \citet{Johnson_2016}; (8) \citet{Donahue_1992}; (9) \citet{Baliunas_1996}; (10) \citet{Petit_2008}; (11) \citet{Kiraga_2007}; (12) \citet{Gaidos_2000, Mamajek_2008}; (13) \citet{Baliunas_1983}; (14) \citet{Gray_1995}; (15) \citet{Saikia_2016}; (16) \citet{Wright_2004}; (17) \citet{Vaughan_1981, Donahue_1996}; (18) \citet{Donahue_1996, Baliunas_1996}}
\end{deluxetable*}

\begin{acknowledgments}
RW thanks Meridith Joyce, Yaguang Li, Jared Goldberg, and Earl Bellinger each for insightful discussions that improved this manuscript.  
\end{acknowledgments}

\bibliography{veil_cited.bib}{}
\bibliographystyle{aasjournalv7}

\end{document}